\documentclass[12pt]{article}

\usepackage{latexsym}
\usepackage{amssymb}
\usepackage{graphicx}
\usepackage{multibox}
\usepackage{amsmath,amsfonts}
\usepackage{bbm}

\usepackage{slashed}

\topskip \textwidth 480pt

\def\*{\ast}
\def\a{\alpha}

\def\s{\sigma}

\def\x{\xi}

\def\rr{\right}
\def\be{\begin{equation}}
\def\ee{\end{equation}}
\def\bqn{\begin{eqnarray}}
\def\eqn{\end{eqnarray}}

\def\theequation{\thesection.\arabic{equation}}

\def \x{{\tt x}}
\def \y{{\tt y}}
\def\pint{-\!\!\!\!\!\!\!\!\;\int}

\newsavebox{\ver}
\newsavebox{\verp}

\newsavebox{\gorp}
\newsavebox{\toch}

\newcommand{\bee}{\begin{eqnarray}}
\newcommand{\eee}{\end{eqnarray}}

\newcommand{\rf}[1]{(\ref{#1})}
\def \foot {\footnote}
\def \no {\nonumber}
\def \ov {\over} \def \rR {{R}}

\def \rr {{\rm r}}

\def \psu {$\mathfrak{psu}(1,1|2)$\ }

\newcommand{\R}{\mathrm{Re}\,}
\newcommand{\I}{\mathrm{Im}\,}

\date{}
\begin{document}
\begin{titlepage}
\title{
\begin{flushright}
{\small MIFPA-11-11}\\
\vskip -0.4cm
{\small NORDITA-2011-30}\\
\vskip -0.4cm
{\small Imperial-TP-AT-2011-2}\\
~\\
\end{flushright}
{\bf Superstrings in $AdS_2\times S^2\times T^6$}
~\\
\medskip
\medskip
\medskip
\author{Dmitri Sorokin$^a$, Arkady  Tseytlin$^b$\footnote{Also at Lebedev Institute, Moscow.}, Linus Wulff$^c$ and Konstantin Zarembo$^{d}$\footnote{Also at ITEP, Moscow, Russia}
~\\
~\\
{\small $^a$\it INFN, Sezione di Padova, via F. Marzolo 8, 35131 Padova, Italia}
~\\
~\\
{\small$^b$\it Blackett Laboratory, Imperial College, London SW7 2AZ, U.K.}
~\\
~\\
\small{$^c$\it George and Cynthia Woods Mitchell Institute}
~\\
{\small\it for Fundamental Physics and Astronomy,}
~\\
{\small\it Texas A\&M University, College Station, TX 77843, USA}
~\\
~\\
{\small$^d$\it Nordita, Roslagstullsbacken 23, SE-106 91 Stockholm, Sweden}
}
}
\maketitle

\begin{abstract}
We consider  the type IIB  Green-Schwarz superstring theory
on  $AdS_2\times S^2\times T^6$  supported  by  homogeneous   Ramond--Ramond
5--form flux and its type IIA T--duals. One  motivation is  to understand the solution of this theory based on
integrability.
This  background is  a limit  of a 1/4
supersymmetric supergravity solution describing four intersecting
D3--branes and represents a  consistent embedding of $AdS_2\times S^2$
into critical  superstring theory.
Its   $AdS_2\times S^2$  part with corresponding
 fermions  can be described by  a classically integrable
  $ PSU(1,1|2)/SO(1,1) \times U(1)$ supercoset sigma--model.
 We point out that since  the RR 5--form  field
 has non--zero components along the 6--torus directions
one cannot, in general, factorize  the  10d  superstring theory
into the   supercoset part  plus 6 bosons and 6 additional massless fermions.
Still, we demonstrate that  the  full superstring model
(i)  is classically integrable, at least
to quadratic order in fermions,  and (ii) admits a consistent classical  truncation to the
 supercoset  part.
Following the analogy with other integrable backgrounds and starting with the
finite-gap equations of the  $ PSU(1,1|2)/SO(1,1) \times U(1)$  supercoset
we propose a set of asymptotic Bethe ansatz equations for a
subset of the quantum string states.

\end{abstract}

\thispagestyle{empty}
\end{titlepage}

\section{Introduction}
Recent remarkable progress in exact solution of the maximally supersymmetric
case of AdS/CFT duality  (see, e.g., \cite{Beisert:2010jr})  suggests
applying similar integrability-based methods to other
 AdS/CFT systems.
We are going to concentrate on the  $AdS_2 \times S^2$ background, whose string sigma model, we will argue, is completely integrable.
The  role   of  $AdS_2 \times S^2$  as the near--horizon geometry of  extremal 4d
Reissner--Nordstr\"om black holes emphasizes the importance of
understanding the corresponding AdS$_2$/CFT$_1$ duality \cite{Maldacena:1997re}.
There is a long and still  unresolved controversy  about the meaning of the corresponding
``CFT$_1$'' (a large $N$ superconformal quantum--mechanical system or a chiral
``half'' of a  2d CFT) \cite{Maldacena:1997re,Gibbons:1998fa,Strominger:1998yg,Maldacena:1998uz,Michelson:1999zf}.
One may hope to shed light on this issue by starting from  the AdS$_2$  side,
solving the corresponding string theory for any value of the radius or effective
tension and re--interpreting the solution  in terms of some  dual  CFT.

The first step is to embed the $AdS_2 \times S^2$ background into critical  10d
superstring theory. Requiring that the bosonic  part of the  string sigma model
should be exactly $AdS_2 \times S^2\times T^6$ excludes
 embeddings with NS--NS flux.\footnote{One may
formally embed the 4d extremal  RN black hole into 10d  string theory using a 6d bosonic
 NS--NS
background  which reduces in the near--horizon limit to (an orbifold of)
an $SL(2,R) \times SU(2)$ WZW  model \cite{Cvetic:1995bj}  but in this case
 the $AdS_2 \times S^2$ coordinates will be coupled to 2 extra compact bosonic
 coordinates.
The same applies  to  various $T$--dual backgrounds \cite{Tseytlin:1996bh}  and  to
 similar heterotic string embeddings of $AdS_2 \times S^2$
\cite{Israel:2004vv,Orlando:2005vt}.}
The relevant RR--flux  embedding into type IIB  string theory
is based on the 1/4 supersymmetric  background describing four intersecting D3--branes \cite{Klebanov:1996mh}.
Its ``near--horizon'' limit  is   $AdS_2 \times S^2\times T^6$
supported, like in the $AdS_5 \times S^5$ case,
by a  homogeneous self--dual 5--form flux. One  may also consider a T--dual
type IIA  background, \emph{e.g.}, the one  based on a superposition of three
 D4--branes and one D0--brane; in that case
the  $AdS_2 \times S^2\times T^6$ space is
supported by a combination of 4--form and 2--form fluxes
but  should lead to an equivalent string theory.\footnote{There
 are also other T-dual cases, e.g. a
D4D4D2D2  configuration  with M5M5M2M2  as its  11d supergravity
 lift \cite{Tseytlin:1996bh,Klebanov:1996mh}
(see also \cite{Duff:1998us,Boonstra:1998yu,Lee:1999yu}).
One may also consider 10d embeddings  of $AdS_2 \times S^2$ with  $T^6$ replaced by a
Calabi--Yau space \cite{Maldacena:1997de,Simons:2004nm}.
}
The spectrum of the corresponding BPS supergravity fluctuation modes was discussed
 in
\cite{Michelson:1999kn,Corley:1999uz,Lee:1999yu,Lee:2000ur}.

A natural  framework for
a superstring theory on a RR background such as $AdS_2\times S^2\times T^6$ is the Green-Schwarz (GS) formalism. The GS action is in principle defined on any supergravity
 background \cite{Grisaru:1985fv}, although constructing it explicitly, in general, is a technically complicated problem.  One needs to know
 the exact form of all background superfields, which
can be reconstructed from the bosonic fields
by solving the supergravity constraints order by order in fermions. The expressions quickly become complicated making this direct approach impractical in the absence of extra symmetries.
 In the $AdS_5 \times S^5$   case  these difficulties were  effectively
 bypassed \cite{Metsaev:1998it}  by
 observing that the  GS action is equivalent to a
supercoset  sigma model on $ PSU(2,2|4)/SO(1,4) \times SO(5)$.

Following the analogy with the construction in
\cite{Metsaev:1998it} and taking into account that the superisometries of the $AdS_2
 \times S^2$   background
form the $PSU(1,1|2)$ supergroup,
ref. \cite{Zhou:1999sm}  found   a formal 4d  GS superstring  action
for the supercoset
$ PSU(1,1|2)/SO(1,1) \times U(1)$.
 A corresponding worldsheet
N=2 superconformal analog  based on the ``standard'' supercoset form of the
action (with a quadratic kinetic term for the fermionic current
 which is absent from the GS action) was constructed in
 \cite{Berkovits:1999zq} where the $\mathbb Z_4$-structure of the
 supercoset and a  local form of the Wess--Zumino
  term were pointed out. In addition to the supercoset part, the action of
   \cite{Berkovits:1999zq} included an $N=2$
   superconformal theory on $CY_3$ (or $T^6$) with 6 worldsheet fermions,
   which is  completely decoupled from the supercoset sigma--model. As
   we shall comment later, the relation of this ``hybrid" model to the
    $AdS_2\times S^2\times T^6$  GS
    superstring remains an open issue.

The $PSU(1,1|2)/SO(1,1) \times U(1)$ supercoset sigma--model
 was interpreted in \cite{Zhou:1999sm}  as a 4d kappa--symmetry invariant GS superstring action in the
 $AdS_2 \times S^2$   background  supported by a  RR 2--form flux.
This  supercoset action  can be viewed as a direct (classical--level)
 truncation of the $AdS_5 \times S^5$
superstring theory with only 4 bosons  and 8 fermions kept non--zero
and has several remarkable features.
The $\mathbb Z_4$--structure \cite{Berkovits:1999zq}  of
the superalgebra \psu  implies \cite{Bena:2003wd}
that this theory is also classically
integrable \cite{Adam:2007ws} and, in addition, self--dual under the fermionic T-duality \cite{Adam:2009kt,Dekel:2011qw}. In fact, its classical integrable structure is essentially
equivalent to that
of the $N=2$ supersymmetric sine--Gordon theory \cite{Kobayashi:1991st} as
the latter is its Pohlmeyer reduction \cite{Grigoriev:2007bu}.

As for the quantum level,  the  rigid symmetry structure  of this
supercoset  GS sigma--model implies that it  should be UV  finite, like its hybrid cousin in \cite{Berkovits:1999zq}, and thus
define a 2d conformal theory. Its one-loop
beta-function was indeed shown to vanish \cite{Zarembo:2010sg} due to the vanishing
Killing form of
the $\frak{psu}(1,1|2)$ superalgebra.
There is, however, an obvious  problem  with interpreting the coset sigma-model as a critical string theory.
The standard  central charge count in a critical flat--space
GS theory  is (see, e.g.,  \cite{Drukker:2000ep}):
$c_{tot} = (n_b +2) - 26 + { \frac{1}{2} } \times 4 \times  n_f$
where $n_b$ and $n_f$ are the numbers of physical bosonic and fermionic degrees of freedom.
While for the 10d  superstring one has
$n_b=n_f=8$ implying $c_{tot} =0$ for the  4d GS  string  one gets
 $c_{tot} =  (2 +2) - 26 + { \frac{1}{2} } \times 4 \times  2 = -18$.
One may try  to cancel  the central charge  deficit by adding  extra decoupled bosons
and fermions but while this may be  straightforward in the NSR framework
it is  not clear a priori  how this can be   consistently implemented  in the GS case.
 An   alternative is to use the ``hybrid'' model of \cite{Berkovits:1999zq} but as mentioned its
equivalence to the  critical 10d  superstring theory remains an open question.

\

In this paper we propose to start directly with a critical 10d  superstring theory defined
in the $AdS_2 \times S^2\times T^6$  background supported by RR flux
and explore its relation to the above supercoset theory.
Since,\emph{ e.g.},  in the type IIB embedding  \cite{Klebanov:1996mh} the $F_5$ flux  has  non--zero components
along the $T^6$ directions\footnote{The  $T^6$  components of the corresponding
 energy--momentum tensor of $F_5$ are
of course equal to zero.}
 the  toroidal string coordinates
do not \emph{a priori } decouple from the GS fermions.
This non-decoupling of the ``flat directions" sets the $AdS_2 \times S^2\times T^6$
 case apart from the previously studied coset-type critical-string backgrounds, where
 additional degrees of freedom could be either completely eliminated ($AdS_4\times CP^3$) or decoupled ($AdS_3\times S^3\times T^4$) from the
 coset by an appropriate choice of kappa-symmetry gauge.
Contrary to what one might expect, in the present case
it will not be possible to represent  the world-sheet sigma-model as
a direct sum of the  $PSU(1,1|2)/SO(1,1) \times U(1)$ supercoset and
additional  free bosonic and fermionic modes.
However, as we will show later,
the GS action  admits  a reduction
to  the supercoset theory in a weaker sense as a classically consistent
truncation.


The mixing between the coset and flat directions of $T^6$ may cast doubts on the potential integrability of the model, beyond the coset truncation.
We shall find that quite remarkably the full
superstring theory in the supersymmetric $AdS_2 \times S^2\times T^6$ backgrounds is also classically integrable.
 Following
\cite{Sorokin:2010wn} we shall construct the Lax
connection (to quadratic order in fermions) from the components of the conserved currents of the full GS action. We will show that the flatness condition for this Lax connection is equivalent to the full set of equations of motion of the GS superstring,
 thus proving classical integrability of the string sigma-model.
 If integrability is not spoiled at the
 quantum level, the theory may
 eventually
be solvable by Bethe ansatz techniques.

As a first step towards the exact solution, we derive the classical counterpart of the Bethe equations that describe   finite--gap classical solutions of the supercoset sigma--model.
Under the assumption that the supercoset truncation
goes through also at the level of massive quantum states with large quantum numbers
we propose
 a set of asymptotic Bethe equations for part of the quantum spectrum of the theory.
We shall also discuss   some preliminary consistency checks of integrability based on
a 1-loop semiclassical expansion.

This paper is organized as follows.
In Section 2  we shall start  with  the action of a  $D=10$ GS superstring in a
general symmetric--space background supported by
 RR fluxes written to quadratic order in fermions
 and find the conditions  for its classical integrability to this order
 by constructing a candidate Lax connection in terms of the Noether isometry currents.

In Section 3 we shall consider explicit examples  of type IIA and type IIB
backgrounds with $AdS_2 \times S^2\times T^6$ supported by RR fluxes and  verify that
the conditions for classical integrability of the corresponding GS superstring action are satisfied,
at least  to quadratic  order in fermions.

In Section 4 we show  that  the type II GS string action
for the  $AdS_2\times S^2\times T^6$ backgrounds
preserving eight supersymmetries can be consistently truncated, at the classical level,
 to the $PSU(1,1|2)/SO(1,1)\times U(1)$ supercoset GS sigma--model with
 $AdS_2\times S^2$ as bosonic part and  eight fermionic modes.
 We shall first  show this  to quadratic order in fermions and then extend the
 argument to all orders.

In Section 5 we shall  consider   two inequivalent  BMN  limits of the superstring action:
When the center of mass (c.o.m.) of the
string is moving along a big circle of $S^2$ or  when it moves both in $S^2$  and in $S^1
\subset  T^6$. The first case corresponds to the BPS vacuum state of the theory and preserves 1/2 of the original supersymmetry. The unbroken supersymmetry can be recast in the 2d form, and leads to the Bose-Fermi degeneracy of the BMN modes. The non-coset degrees of freedom remain massless in this limit.
In the second case the gauge--fixed sigma--model does not have effective 2d supersymmetry, and the degeneracy is lifted. Moreover, some of the `non--supercoset' fermions acquire mass in this case.

In Section 6 we shall first review the classical Bethe equations describing
finite gap solutions  of  the $PSU(1,1|2)/SO(1,1)\times U(1)$
 supercoset model \cite{Zarembo:2010yz}
 and then propose a set of quantum  asymptotic Bethe equations that potentially describe the spectrum of the string in the light-cone gauge associated with the supersymmetric BMN geodesic. These equations, however, only describe
 a subset of the massive states and do not capture the massless, non-coset degrees of freedom.

 In Section 7 we  discuss  various  semiclassical string solutions in
 $AdS_2\times S^2\times T^6$ and clarify the validity of the
  assumption of decoupling of
 the $AdS_2\times S^2$ sector.

There are several Appendices describing our notation, the structure  of the (enlarged) \psu
superalgebra and  properties of Killing vectors  on symmetric spaces.
In Appendix D we also  provide details of
the derivation of the classical string equations of motion
from the flatness of the Lax connection constructed in Section 2.

\section{GS superstrings in  RR  backgrounds and their\\
 classical integrability}

A particular feature of GS superstrings
propagating in supersymmetric backgrounds whose geometry is a direct product of an AdS
 space with compact manifolds is that (at least part of) their dynamics can often be described
 by certain  supercoset sigma--models whose isometry superalgebras have a $\mathbb
  Z_4$--grading.
  For example,  the  $PSU(2,2|4)/SO(1,4)\times SO(5)$ sigma--model  defines  the
   maximally supersymmetric type IIB $AdS_5\times S^5$ superstring theory
    \cite{Metsaev:1998it}. Also,
     an $OSp(6|4)/SO(1,3)\times U(3)$ sigma--model with  24 supersymmetries
    \cite{Arutyunov:2008if,Stefanski:2008ik}
    represents  the  type IIA $AdS_4\times
    CP^3$ superstring in those sectors where
     a partial kappa--symmetry gauge
     fixing is allowed to reduce the complete GS  action to the
      $OSp(6|4)/SO(1,3)\times U(3)$ sigma--model one containing    24 worldsheet fermions
    \cite{Arutyunov:2008if,Gomis:2008jt,Grassi:2009yj,Cagnazzo:2009zh}.

In the case of an $AdS_3\times S^3\times T^4$ background supported by RR 3-form flux
which preserves 16 out of the 32 possible
 supersymmetries\footnote{This background can be regarded as a particular limit
	of an $AdS_3\times S^3 \times S^3 \times S^1$ background in which one $S^3$  is
	``re--decompactified" into $T^3$. The $AdS_3\times S^3 \times S^3 \times S^1$
	backgrounds are associated with the supercosets
$D(1,2;\alpha)\times D(1,2;\alpha)/SO(1,2)\times SO(3)\times SO(3)$, where $D(1,2;\alpha)$
is an  exceptional  supergroup
(see \cite{Babichenko:2009dk} for more details and references).},
 one can reduce the GS
action to a sigma--model on the supercoset $PSU(1,1|2)\times PSU(1,1|2)/SU(1,1)\times SU(2)$
(which has  $AdS_3\times S^3$ as its bosonic subspace)
 plus the decoupled free bosonic sector
 on $T^4$ \cite{Babichenko:2009dk}.
 To this end one should completely gauge fix the
 kappa--symmetry by eliminating 16 of the 32  fermionic modes
  of the  10d  GS superstring in an appropriate way.
  For some ``singular''
   classical string configurations (e.g.,  when the string does not
   wind or move in $T^4$) such kappa--symmetry gauge fixing is  inadmissible.
   In such special  cases  there are physical fermionic degrees of freedom which are not part of the
   supercoset sigma--model  and to take them into account one should
   start with the  complete 10d GS    superstring action.

The situation turns out to be   more complicated in the case of
$AdS_2\times S^2\times T^6$ backgrounds.
 The GS sigma--model on  $PSU(1,1|2)/SO(1,1)\times U(1)={\rm Super}(AdS_2\times S^2)$
\cite{Zhou:1999sm} indeed possesses a $\mathbb Z_4$--grading \cite{Berkovits:1999zq} and is thus
classically integrable \cite{Bena:2003wd}, much like the GS sigma-models on $AdS_5\times S^5$, $AdS_4\times CP^3$ and $AdS_3\times S^3\times T^4$. The crucial difference is in the number of preserved supersymmetries. The $\frak{psu}(1,1|2)$ superalgebra has only 8 supercharges,
 compared to 32 for $AdS_5\times S^5$, 24 for
$AdS_4\times CP^3$ and 16 for $AdS_3\times S^3\times T^4$.
As a result,
 the 16-parametric kappa--symmetry of the
GS action is not sufficient to gauge away the additional $32-8=24$ worldsheet
fermions  associated with broken
 10d supersymmetries.
 At least 8 of the broken-symmetry fermions will remain in the physical spectrum of the string, and will link the decoupled, at first sight, $T^6$ sector to the coset.

Since extra, non-coset degrees of freedom cannot be gauged away and do not decouple, the integrability of the supercoset is not sufficient to prove the integrability of the full
 theory which includes the fermions associated with  broken supersymmetries.
  To analyze integrability, we will follow an alternative
   approach, proposed for the type IIA superstring action
    on $AdS_4\times CP^3$ \cite{Sorokin:2010wn}.
  In this approach, the Lax connection is constructed directly from the Noether currents of the full superstring action without requiring that the string sigma--model has a coset structure. We will
  demonstrate that (at least up to the second order
 in all the 32 fermions) the type II $D=10$ GS superstring  propagating in $AdS_2\times
  S^2\times T^6$ background  (supported by type IIA or type IIB RR  fluxes)
  is classically   integrable.
   Thus the integrability  still applies despite the fact
    that  the integrable
   $PSU(1,1|2)/SO(1,1)\times U(1)$ sigma--model cannot give
    a complete description of the   GS superstring in the $AdS_2\times
  S^2\times T^6$  case.

We shall also show that in the non--supersymmetric $AdS_2\times S^2\times T^6$ backgrounds
 which are
obtained from the supersymmetric ones by changing signs of some components of the
RR fluxes, the same prescription for the construction of the Lax connection
fails already at the quadratic order in fermions.

\

In  this section we will consider a $D=10$ GS superstring in a generic
superbackground with non--zero RR fluxes\footnote{The analysis of the
integrability of GS superstrings in backgrounds with NS--NS fluxes turns out to
be more complicated and requires a separate consideration, in particular because
 in these cases the purely bosonic NS--NS 2--form contributes to the WZ term of
 the string action.} whose bosonic subspace is a symmetric space. We  will derive
  relations that should be satisfied by components of the Noether currents of
 the background superisometries in order to be able to construct from them a zero--curvature Lax
    connection, up to the second order in fermions.
    We then show that the type II superstrings  in the $AdS_2\times S^2\times T^6$  backgrounds
     satisfy these
    requirements and hence are classically integrable, at least to
    quadratic order in fermions.

To construct a complete Lax connection to all orders in fermions one should know
  the explicit form of the  full  GS superstring action.
 A  derivation of such an explicit
 form of the GS action in a $D=10$ superbackground which is {\it not} maximally supersymmetric turns
  out to be a technically complicated problem. So far it has been solved only for two
   cases which can be obtained by double dimensional reduction of the
   supermembrane action in corresponding $D=11$  maximally supersymmetric
   backgrounds: a locally
   flat one and $AdS_4\times S^7$. They correspond to, respectively,
   the type IIA superstring in a 7--brane
    background with a magnetic RR flux \cite{Russo:1998xv} and to the  type IIA superstring in
    the $AdS_4\times CP^3$
    superbackground \cite{Gomis:2008jt} (the latter is the extension to superspace of
     the Hopf fibration realization of the $D=11$ supergravity solution
     \cite{Nilsson:1984bj,Sorokin:1984ca,Sorokin:1985ap}).

\subsection{GS superstrings in RR backgrounds, equations of motion
and conserved currents}
The action for the GS superstring in a bosonic supergravity background
(with zero  NS--NS flux and constant dilaton $\phi$) has the
following form up to quadratic
order in fermions \cite{Tseytlin:1996hs,Cvetic:1999zs}
\footnote{In what follows we shall mainly use the language of 2d differential forms with wedge products understood.}
\begin{equation}\label{action}
S=-T\int\left(\frac{1}{2}\ast e^Ae_A+i\ast
e^A\,\Theta\Gamma_A{\mathcal D}\Theta-ie^A\,\Theta\Gamma_A\hat\Gamma\, {\mathcal D}\Theta\right)\,,
\end{equation}
where $e^A(X)$ $(A=0,1,\cdots,9)$ are worldsheet pull--backs
of the background vielbein one--forms
and
\begin{equation}\label{E}
{\mathcal D}\Theta=(\nabla-\frac{1}{8}e^A\,\slashed F\Gamma_A)\Theta \,,
\end{equation}
is the fermionic vielbein to the lowest order in
fermions $\Theta$. Here
$\nabla=d+\omega$ is the covariant derivative containing the spin connection of
the background space--time,
\begin{equation}
\hat\Gamma=\left\{
\begin{array}{c}
\Gamma_{11}\\
\sigma^3
\end{array}
\right.
\begin{array}{c}
(\mathrm{IIA})\\
(\mathrm{IIB})
\end{array}
\end{equation}
and the coupling to the RR fields is given in terms of the matrix
\begin{equation}
\label{eq:slashedF}
\slashed F=
e^\phi\left\{
\begin{array}{c}
-\frac{1}{2}\Gamma^{AB}\Gamma_{11}F_{AB}+\frac{1}{4!}\Gamma^{ABCD}F_{ABCD}\\
{}\\
i\sigma^2\Gamma^AF_A-\frac{1}{3!}\sigma^1\Gamma^{ABC}F_{ABC}+\frac{i}{2\cdot5!}\sigma^2\Gamma^{ABCDE}F_{ABCDE}
\end{array}
\right.
\begin{array}{c}
(\mathrm{IIA})\\
{}\\
(\mathrm{IIB})
\end{array}
\end{equation}
in the type IIA and type IIB case, respectively\footnote{Our definition of $\Gamma_{11}$ and type IIA RR--fluxes differs by a sign from those of \cite{Cvetic:1999zs}.}. We describe the two
Majorana--Weyl spinors in the IIA case as one 32-component Majorana spinor
$\Theta$ and in the IIB case as two 32-component Majorana spinors projected onto
one chirality $\Theta^i$ ($i=1,2$) by
$\frac{1}{2}(1+\Gamma_{11})$.
 The Pauli matrices
$\sigma^1,\sigma^2$ and $\sigma^3$ act on the IIB SO(2)--indices $i,j=1,2$
which will be suppressed.

The bosonic equations of motion  to the second order in $\Theta$ are then
\bee\label{bosoneq}
&\nabla\left(*(e^A+i\Theta\Gamma^A {\mathcal D}\Theta)+i\Theta\Gamma^A\hat\Gamma\,{\mathcal D}\Theta
-\frac{i}{8}*e^B\,\Theta\Gamma^A\slashed F\Gamma_B\Theta
-\frac{i}{8}e^B\,\Theta\Gamma^A\hat\Gamma\slashed F\Gamma_B\Theta\right)
 &\\
&
-\frac{i}{4}*e^Be^E\,\Theta\Gamma_B{}^{CD}\Theta\,R_{CDE}{}^A
+\frac{i}{4}e^Be^E\,\Theta\Gamma_B{}^{CD}\hat\Gamma\Theta\,R_{CDE}{}^A=0\,,&\nonumber
\eee
where $R_{CDE}{}^A(X)$ is the curvature of the $D=10$ space, and the fermionic
equations (linear in $\Theta$) are
\begin{equation}\label{fe}
(\ast e^A\,\Gamma_A-e^A\,\Gamma_A\hat\Gamma)\,{\mathcal D}\Theta=0\,.
\end{equation}
If the background has bosonic isometries, generated by Killing vectors
$K_A(X)$, the worldsheet model has the  corresponding conserved Noether current one--form
of the  following generic form (see \cite{Sorokin:2010wn} for more details)
\begin{equation}\label{BC}
J_{\mathcal B}=J^AK_A+J^{AB}\nabla_AK_B=e^AK_A+\mbox{fermions}\,,
\end{equation}
where the second term comes from compensating Lorentz transformations of the
fermionic fields $\Theta$.
 The $J^A$ and $J^{AB}$ terms in the current have the following form
\begin{equation}\label{JA}
J^A=e^A+i\Theta\Gamma^A {\mathcal D}\Theta-\frac{i}{8}e^B\,\Theta\Gamma^A\slashed F\Gamma_B\Theta+i\Theta\Gamma^A\hat\Gamma\*\mathcal D\Theta
-\frac{i}{8}*e^B\,\Theta\Gamma^A\hat\Gamma\slashed F\Gamma_B\Theta\,,
\end{equation}
\begin{equation}\label{JAB}
J^{AB}=
\frac{1}{4}(\Gamma^{AB}\Theta)^\alpha\ast i_\alpha\mathcal L
=
-\frac{i}{4}e^C\,\Theta\Gamma^{AB}{}_C\Theta
+\frac{i}{4}\ast e^C\,\Theta\Gamma^{AB}{}_C\hat\Gamma\Theta\,,
\end{equation}
where $\mathcal L$ is the superstring Lagrangian in \eqref{action}.

If the background preserves some supersymmetries generated by Killing
spinors $\Xi(X)$, there is also a conserved supersymmetry current on the
worldsheet which to linear order in $\Theta$ has the following form
\begin{equation}\label{susy}
J_{susy}=\frac{i}{2R}(e^A\,\Theta\Gamma_A\Xi-\ast
e^A\,\Theta\Gamma_A\hat\Gamma\Xi)\,,
\end{equation}
where the dimension--of--length constant $R$ (which in our case will be the AdS radius) has been introduced
to make the current dimensionless, and $\Xi$ satisfies the Killing spinor equation
\begin{equation}
\nabla\Xi-\frac{1}{8}e^A\,\slashed F\Gamma_A\Xi=0\,.
\end{equation}
As usual, the  currents \eqref{BC} and \eqref{susy} are conserved
\be\label{d*J}
d*J_{\mathcal B}=0,\qquad d*J_{susy}=0,
\ee
provided that the equations of motion \eqref{bosoneq} and \eqref{fe} are
satisfied (and vice versa).
The conservation of \eqref{BC} and orthogonality of $K_A$ and
$\nabla_AK_B=[K_A,K_B]$ imply that the following relations must hold separately (see Appendix C for basic relations satisfied by the Killing vectors of a symmetric space)
\begin{eqnarray}
(\nabla\ast J^{AB}-\ast J^{[A}e^{B]})K_A\,K_B=0\,,\label{eq:nablaJAB}\\
\nabla\ast J^A-2R_{BCD}{}^A\ast J^{CD}e^B=0\,.
\end{eqnarray}

\subsection{Lax connection built from conserved currents}
In many examples of interest, such as $AdS_4\times CP^3$, $AdS_2\times S^2\times T^6$
or $AdS_3\times S^3\times T^4$, the complete target superspace describing the
superstring background is not a supercoset manifold, so the
$\mathbb Z_4$--graded supercoset prescription of ref. \cite{Bena:2003wd} for the construction of the Lax connection does not directly apply. In \cite{Sorokin:2010wn}
an alternative prescription was proposed for these cases and applied to
$AdS_4\times CP^3$. It uses   the conserved currents as building
blocks. The Lax connection has two parts \footnote{When reduced to the supercoset sigma--model this alternative Lax connection is related to the conventional one by a superisometry gauge transformation \cite{Sorokin:2010wn}.}
\begin{equation}\label{Lax}
L=L_{\mathcal B}+L_{\mathcal F}\,.
\end{equation}
The part $L_{\mathcal B}$ of the Lax connection that corresponds to the bosonic
isometries has the following form (similar to \eqref{BC})
\begin{equation}\label{LB}
L_{\mathcal B}=L^AK_A+L^{AB}\nabla_AK_B\,,
\end{equation}
where
\begin{eqnarray}
L^A&=&\alpha_1\, e^A+\alpha_2\ast J^A,\\
L^{AB}&=&\alpha_2^2\,J^{AB}+\alpha_2(1+\alpha_1)\ast J^{AB}\,.
\end{eqnarray}
The part  corresponding to the fermionic isometries is
\begin{equation}\label{LF}
L_{\mathcal F}=-\alpha_2\beta_1J_{susy}+\alpha_2\beta_2\ast J_{susy}\,.
\end{equation}
The numerical parameters $\alpha_1$, $\alpha_2$, $\beta_1$ and $\beta_2$  are expressed in terms of a single spectral parameter by
requiring that the Lax connection \eqref{Lax} has zero curvature
\be\label{zeroR}
dL-LL=0\,.
\ee
In fact, the values of $\alpha_1$ and $\alpha_2$ are determined in terms of the spectral parameter already by the requirement of integrability of the
purely bosonic part of the superstring sigma--model. One can check that if in
\eqref{LB} we put $\Theta$ to zero, the Lax connection reduces to its
conventional form for an integrable sigma--model on a  symmetric space
\cite{Eichenherr:1979ci}:
$$
L_{\mathcal B}|_{\Theta=0}=(\alpha_1e^A+\alpha_2\ast e^A)K_A\,.
$$
The curvature of this connection vanishes if
\begin{equation}\label{spectral}
\alpha_2^2=2\alpha_1+\alpha_1^2\, \qquad \Rightarrow \qquad
\alpha_1=\frac{2{\tt x}^2}{1-{\tt x}^2}, \qquad \alpha_2=\pm\frac{2{\tt x}}{1-{\tt x}^2},
\end{equation}
where ${\tt x}$ is the spectral parameter.
When the fermionic fields are non--zero, using the equations of motion we find
that up to the quadratic order in fermions the $L_{\mathcal B}$--terms in the curvature
\eqref{zeroR} proportional to the Killing vectors $K_A$ are
\begin{eqnarray}\label{KA}
(dL_{\mathcal B}-L_{\mathcal B}L_{\mathcal
B})^A&=&2\alpha_2(\alpha_2^2-2\alpha_1-\alpha_1^2)\,R_{BCD}{}^A\ast J^{CD}e^B\,.
\end{eqnarray}
We observe that the right-hand-side of \eqref{KA} vanishes when the parameters
satisfy eq. \eqref{spectral}.
The $L_{\mathcal B}$--terms of the Lax curvature proportional to the derivative
of the Killing vectors $\nabla_AK_B$ are
\begin{equation}
\label{eq:dLAB}
(dL_{\mathcal B}-L_{\mathcal B}L_{\mathcal B})^{AB}=\alpha_2^2\Big[\nabla
J^{AB}+(J-e)^{[A}e^{B]}\Big]\,.
\end{equation}
If the right-hand-side of \eqref{eq:dLAB} vanishes then $L_{\mathcal B}$ has
zero curvature and may itself play the role of the Lax connection (i.e. in \eqref{Lax} the term $L_{\mathcal F}$ can be set to zero). An example, when this happens is
\emph{an integrable non--supersymmetric} GS superstring on $AdS_4$ considered in
\cite{Sorokin:2010wn}. In this case there is no supersymmetry Noether current \eqref{susy}.

 However, in general, the right-hand-side of \eqref{eq:dLAB} is
non--zero. Then we need to show that these terms in the curvature can be cancelled
by adding other terms to the Lax connection. When the background possesses
some supersymmetry these additional terms are built from the
supersymmetry current \eqref{susy} and have the form given in \eqref{LF}. This is what happened in $AdS_4\times
CP^3$ \cite{Sorokin:2010wn}. It will be possible to cancel the curvature of the
 Lax connection by this mechanism for  the
$AdS_2\times S^2\times T^6$ backgrounds as well.

To see if the curvature term \eqref{eq:dLAB} can be canceled by contributions
from \eqref{LF}, let us compute the curvature of the total Lax connection \eqref{Lax} for the parameters $\alpha_1$ and $\alpha_2$ related by \eqref{spectral}
\begin{eqnarray}\label{R}
dL-LL&=&
\alpha_2^2(\nabla J^{AB}+(J^A-e^A)e^B)\nabla_AK_B
+\alpha_2^2(\beta_2^2-\beta_1^2)J_{susy}^2
-\alpha_2\beta_1dJ_{susy}
\nonumber\\
&&{}
+\alpha_2(\alpha_1\beta_1+\alpha_2\beta_2)(J_{\mathcal
B}J_{susy}+J_{susy}J_{\mathcal B})
\nonumber\\
&&{}
-\alpha_2(\alpha_1\beta_2+\alpha_2\beta_1)(J_{\mathcal B}\ast J_{susy}+\ast
J_{susy}J_{\mathcal B})\,.
\end{eqnarray}
Now, if
\begin{eqnarray}\label{coef}
\beta_1=\mp\sqrt{\frac{\alpha_1}{2}}=\pm\frac{i{\tt x}}{\sqrt{{\tt x}^2-1}}\,,\qquad
\beta_2=\pm\frac{\alpha_2}{\sqrt{2\alpha_1}}=\mp\frac{i}{\sqrt{{\tt x}^2-1}}\,
\end{eqnarray}
 the last term in eq. \eqref{R} vanishes and the other terms cancel
each other provided that
\begin{equation}
dJ_{susy}=-2(J_{\mathcal B}J_{susy}+J_{susy}J_{\mathcal B}),
\label{eq:inteq2}
\end{equation}
and
\begin{equation}
\left[\nabla J^{AB}+(J^A-e^A)e^B\right]\nabla_AK_B=-J_{susy}^2\,.
\label{eq:inteq1}
\end{equation}
These equations can be viewed as a modified version of the familiar Maurer-Cartan equation $dJ=-2J^2$, with $J=J_{\mathcal B}+\frac{1}{\sqrt2}J_{susy}$ and with some terms proportional to $K_A$ removed.

Verifying that eqs. (\ref{eq:inteq2}) and (\ref{eq:inteq1})  hold is therefore
enough to demonstrate the integrability at this order. We will see below that just
as in the $AdS_4\times CP^3$ case these equations hold for strings in
$AdS_2\times S^2\times T^6$.
 Let us first compute the left-hand-sides of
these equations in a general RR background.
Using the form of the supersymmetry current \eqref{susy} and the equations of
motion we get
\begin{equation}
\label{eq:dJSUSY}
dJ_{susy}=\frac{i}{8R}(
e^Ae^B\,\Theta\Gamma_A\slashed F\Gamma_B\Xi
-\ast e^Ae^B\,\Theta\Gamma_A\hat\Gamma\slashed F\Gamma_B\Xi)\,.
\end{equation}
Using the form \eqref{JAB} of $J^{AB}$
we find that
\begin{eqnarray}
\nabla J^{AB}&=&
-\frac{i}{2}e^C\,\Theta\Gamma^{AB}{}_C\nabla\Theta
+\frac{i}{2}\ast e^C\,\Theta\Gamma^{AB}{}_C\hat\Gamma\nabla\Theta
\nonumber\\
&=&
ie^{[B}\,\Theta\Gamma^{A]}\nabla\Theta
-i\ast e^{[B}\,\Theta\Gamma^{A]}\hat\Gamma\nabla\Theta
-\frac{i}{16}e^Ce^D\,\Theta\Gamma^{AB}\Gamma_C\slashed F\Gamma_D\Theta
\nonumber\\
&&{}
+\frac{i}{16}\ast e^Ce^D\,\Theta\Gamma^{AB}\Gamma_C\hat\Gamma\slashed
F\Gamma_D\Theta\,,
\end{eqnarray}
where we have made use of the equations of motion. We can further rewrite this as
\be\label{j-ej}
\nabla J^{AB}+e^{[A}(J^{B]}-e^{B]})
=
-\frac{i}{16}e^Ce^D\,\Theta\Gamma_C\Gamma^{AB}\slashed F\Gamma_D\Theta
+\frac{i}{16}\ast e^Ce^D\,\Theta\Gamma_C\Gamma^{AB}\hat\Gamma\slashed
F\Gamma_D\Theta\,.
\ee
As we have seen, for the Lax--connection curvature \eqref{R} to vanish the
right-hand-side of \eqref{j-ej} should be the square of $J_{susy}$ and the
right-hand-side of (\ref{eq:dJSUSY}) should be the commutator of $J_{\mathcal
B}$ with $J_{susy}$. In the next section we will show that this holds for
various $AdS_2\times S^2\times T^6$ solutions with RR fluxes which preserve eight
supersymmetries.


\section{Superstrings in $AdS_2\times S^2\times T^6$ backgrounds}
\setcounter{equation}0

As we have discussed in the Introduction, there are several solutions of type
IIA and type IIB supergravity with the geometry of $AdS_2\times S^2\times T^6$
supported by RR fluxes  which are related by $T$-duality.
Type IIA solutions  are also related  by
 dimensional reduction to certain $AdS_n \times S^k \times T^m $ solutions
of $D=11$ supergravity which correspond to limits of  configurations
of intersecting M2-- and M5--branes (that upon compactification describe
 $D=4$ black holes) \cite{Klebanov:1996mh}.

We start from the most symmetric configuration which consists of four intersecting $D3$-branes in type IIB string theory. The setup is as follows (with $\times$ indicating the directions along the branes)
\begin{equation}
\begin{array}{ll|c|cccccc}
&AdS_2&S^2&&&T^6&&&\\
\hline&0\quad1&2\quad3&4&5&6&7&8&9\\
\hline D3&\times&&\times&\times&\times&&&\\
D3&\times&&\times&&&\times&\times&\\
D3&\times&&&\times&&&\times&\times\\
D3&\times&&&&\times&\times&&\times\\
\hline
\end{array}
\end{equation}
and the  $F_5$--flux is given in \eqref{F5}. By performing T-dualities on some of the $T^6$-directions we can obtain different solutions. In particular, T--dualizing along one of the $T^6$ directions, say $y^4$, gives a configuration with two $D2$ and two $D4$-branes and electric and magnetic $F_4$-flux given in \eqref{F4} in type IIA. Similarly T-dualizing the (456)-directions gives one $D0$ and three $D4$-branes with electric $F_2$-flux and magnetic $F_4$-flux given in \eqref{RR}, while T-dualizing the (789)-directions instead gives one $D6$ and three $D2$-branes with magnetic $F_2$-flux and electric $F_4$-flux given in \eqref{RR1}.

We will consider these solutions in more detail below and verify explicitly the integrability of the GS superstring in the IIA background with electric $F_2$ and magnetic $F_4$--flux as well as in the IIB background with $F_5$--flux to quadratic order in fermions. Since the two backgrounds are related by T--duality it should of course not be necessary to perform this analysis in both cases, however we find this worth doing since there are interesting differences between the two cases in how, for example, the $\mathfrak{psu}(1,1|2)$-algebra is realized. This also gives a useful consistency check of our results.

\subsection{Type IIA $AdS_2\times S^2\times T^6$ with $F_2$ and $F_4$ flux}\label{F2F4}

This 1/4 supersymmetric $AdS_2\times S^2\times T^6$ solution
 of type IIA supergravity is
supported by the following combination of the RR 2-form and 4-form fluxes
\begin{eqnarray}
F_2&=&-\frac{e^{-\phi}}{2R}e^be^a\varepsilon_{ab}\,,\nonumber\\
F_4&=&-\frac{e^{-\phi}}{2R}e^{\hat b}e^{\hat a}\varepsilon_{\hat a\hat b}J_2\,,
\label{RR}
\end{eqnarray}
where $\phi$ is the (constant) dilaton\footnote{We
keep the  explicit dependence of  fluxes on the constant dilaton
to indicate  the dependence on the string coupling constant.},
$J_2=\frac{1}{2}dy^{b'}dy^{a'}J_{a'b'}$ is the K\"ahler form\footnote{
Explicitly, if $z^{\mathcal{I}}, \bar{z}_{\mathcal{I}}$ are holomorphic coordinates on the torus, e.g. $z^1=y^4+iy^5$ etc., we have $J_2=\frac{i}{2}d\bar{z}_{\mathcal{I}}\wedge d{z}^{\mathcal{I}}$
.}
  on $T^6$, $\varepsilon^{01}=1$, $R$ is
the radius of $AdS_2$ and $S^2$, and  $a,b=0,1$, $\hat a,\hat b=2,3$
and $a', b'=4,\ldots,9$ are the $AdS_2$, $S^2$ and $T^6$ indices, respectively.

There is also a solution with the roles of the two fluxes interchanged, i.e.
\begin{eqnarray}\label{RR1}
F_2&=&\frac{e^{-\phi}}{2R}e^{\hat b}e^{\hat a}\varepsilon_{\hat a\hat b}\,,\nonumber\\
F_4&=&-\frac{e^{-\phi}}{2R}e^be^a\varepsilon_{ab}J_2\,.
\end{eqnarray}
This solution can be obtained by reduction from the 1/4 supersymmetric $D=11$ supergravity solution
$AdS_2\times S^3\times T^6$ with the same $F_4$--flux. This is done by
realizing $S^3$ as an $S^1$ Hopf fibration over $S^2$. Dimensionally reducing
on this $S^1$ creates an $F_2$--flux proportional to the K\"ahler form on
$S^2\sim CP^1$ similar to what happens in
the $AdS_4\times CP^3$ case.
 In a similar way,  the type IIA solution \eqref{RR}
 is the dimensional reduction of the 1/4 supersymmetric
$D=11$ supergravity solution $AdS_3\times S^2\times T^6$ with $F_4=-\frac{1}{2R}e^{\hat
b}e^{\hat a}\varepsilon_{\hat a\hat b}J_2$.

Note that if we change the sign of one of the fluxes in \eqref{RR} or \eqref{RR1},
 the $AdS_2\times
S^2\times T^6$ space will still be a solution of the supergravity equations of
motion, but the supersymmetry will be completely broken. This supersymmetry
breaking is reflected in the form of $\slashed F$ in  \eqref{eq:slashedF}
 which will no longer  be proportional to a projector.
This will also  spoil the structure of the Lax connection discussed in the previous section;
 in the non--supersymmetric cases we have not been able to make it have zero
curvature since it does not appear to be possible to cancel the terms
 on the right--hand--side of eq.
\eqref{eq:dLAB} without a supersymmetry current. This may indicate that, though the purely bosonic sector of these non--supersymmetric models is classically integrable, the integrability is spoiled by the fermionic sector.

Let us now consider
 the structure of the superisometry algebra \psu of the
  background \eqref{RR} and demonstrate that
  it ensures the vanishing of the curvature
   of the Lax connection \eqref{Lax} (the
   analysis of the background \eqref{RR1} can be carried out in exactly the same fashion).
For the supersymmetric solution \eqref{RR}   the
definition of $\slashed F$ in eq. (\ref{eq:slashedF}) gives
\begin{eqnarray}\label{FIIA}
&&\slashed F=
-\frac{1}{R}\gamma\Gamma_{11}
+\frac{i}{2R}\gamma\gamma^5\Gamma^{a'b'}J_{a'b'}
=
-\frac{4}{R}\mathcal P_8\gamma\Gamma_{11}\,,\\
&&
\mathcal P_8=\frac{1}{8}(2-iJ_{a'b'}\Gamma^{a'b'}\gamma^7)\,, \label{pii}
\end{eqnarray}
where $\gamma=\Gamma^0\Gamma^1$ is the product of $AdS_2$ gamma matrices (so that $\gamma^2=1$)
 and $\gamma^5=i\Gamma^0\Gamma^1\Gamma^2\Gamma^3$.
Here  $\mathcal P_8$ is a projector   which
for all the $AdS_2\times S^2\times T^6$ solutions   has the
following properties: it commutes with the $D=10$ matrices $\Gamma^a$
$(a=0,1)$ and $\Gamma^{\hat a}$ $(\hat a=2,3)$ along the $AdS_2\times S^2$
directions, as well as with $\hat\Gamma$ ($=\Gamma_{11}$ in the IIA case)
\be\label{PG}
[\mathcal P_8,\Gamma^a]=[\mathcal P_8,\Gamma^{\hat a}]=[\mathcal
P_8,\hat\Gamma]=0\,,
\ee
while for $\Gamma^{a'}$   with $a'=4,\ldots,9$  along the $T^6$ directions we have
\be\label{PTP}
\mathcal P_8\Gamma^{a'}\mathcal P_8=0\,.
\ee
The projector $\mathcal P_8$ singles out an eight--dimensional supersymmetric subspace of the
32 dimensional fermionic space and is defined in a similar way to the $CP^3$ case
\cite{Nilsson:1984bj,Gomis:2008jt} but with the K\"ahler form of $CP^3$
replaced by that of $T^6$.\footnote{For the case of $AdS_4\times CP^3$ we had $\slashed F=-\frac{8i}{R}
(1-\mathcal P_{8})\gamma^5$. Note that in
$AdS_4\times CP^3$ the projector $\mathcal P_{8}$ singled out
eight broken supersymmetry fermions.}

As we shall see, the same projector appears in the \psu
superalgebra of the superisometries and so the extra terms in $\nabla J^{AB}$, eqs. \eqref{eq:inteq1} and \eqref{j-ej}, are of the required form
to be canceled by the terms coming from $J_{susy}^2$.
A conventional form of the \psu  algebra given in eq. (\ref{eq:PSUQ}) of Appendix B differs from
what we would like to have in this case by factors of $\gamma^5$. A way to cure this is to take a
slightly different realization of the gamma matrices in the algebra
(\ref{eq:PSUQ}) which, of course, is an automorphism of the algebra. Taking
\begin{eqnarray}
\Gamma^a&\rightarrow&i\Gamma^a\gamma^5,\nonumber\\
\Gamma^{\hat a}&\rightarrow&i\Gamma^{\hat a}\gamma^5,\nonumber\\
\Gamma^{a'}&\rightarrow&\Gamma^{a'}
\end{eqnarray}
together with the redefinition of the charge conjugation matrix
$\mathcal C\rightarrow i\mathcal C\gamma^5$ preserves the
 Clifford algebra, symmetry properties of the gamma matrices
  and the form of the projector $\mathcal P_8$. The commutators
  involving $Q=\mathcal P_8Q$ in the \psu algebra (\ref{eq:PSUQ}) now become
\begin{eqnarray}
[P_A,Q]&=&\frac{1}{2R}Q\gamma\Gamma_{11}\Gamma_A\mathcal
P_8\ ,\qquad[M_{AB},Q]=-\frac{1}{2}Q\Gamma_{AB}\mathcal P_8
\nonumber\\
\label{repPS112}
\{Q,Q\}&=&
2i(\mathcal P_8\Gamma^A\mathcal P_8)P_A
+\frac{iR}{2}(\mathcal P_8\Gamma^{AB}\gamma\Gamma_{11}\mathcal
P_8)R_{AB}{}^{CD}M_{CD}\,,
\end{eqnarray}
and have the required form since they contain $\slashed F\propto\mathcal
P_8\gamma\Gamma_{11}$.

The relations between the  Killing vectors $K_A$ and the
Killing spinors $\Xi$ and the generators
of the algebra are\footnote{In the following equations we explicitly
 display the charge-conjugation matrices  to avoid possible confusion.}
\begin{eqnarray}
K_A=kP_Ak^{-1},\qquad i\mathcal C\gamma\Gamma_{11}\Xi=kQk^{-1}\\
\nabla_AK_B=-\frac{1}{2}R_{AB}{}^{CD}kM_{CD}k^{-1}\,,
\end{eqnarray}
where $k(x)$ is a coset element of $SO(1,2)\times SU(2)/SO(1,1)\times U(1)$ and the Killing vectors $K_{a'}=P_{a'}$ generate the translation isometries of $T^6$.

 Using these relations we have
\begin{eqnarray}
\{\Xi,\Xi\}&=&
2i(\gamma\mathcal P_8\Gamma^A\mathcal P_8\gamma\mathcal C)K_A
+iR(\mathcal P_8\Gamma^{AB}\gamma\Gamma_{11}\mathcal P_8\mathcal C)\nabla_AK_B
\nonumber\\
{}[K_A,\Xi]&=&-\frac{1}{2R}\Xi\Gamma_A\mathcal P_8\gamma\Gamma_{11}\mathcal C\,.
\end{eqnarray}
 Note that along the $T^6$ directions $\nabla_{a'}K_{b'}=[K_{a'},K_{b'}]=0$,
as should be the case since the $T^6$ translational isometries are abelian.

Using this algebra together with \eqref{j-ej} and \eqref{susy} we get (up to quadratic order in fermions)
\begin{equation}
J_{susy}^2=-\Big[\nabla J^{AB}+e^A(J^B-e^B)\Big]\,\nabla_AK_B
\end{equation}
and, using \eqref{eq:dJSUSY} and \eqref{BC},
\begin{equation}
J_{\mathcal B}J_{susy}+J_{susy}J_{\mathcal B}=-\frac{1}{2}dJ_{susy}\,.
\end{equation}
These equations agree with (\ref{eq:inteq1}) and (\ref{eq:inteq2}). Thus
the GS superstring action
in this background  admits
 the Lax connection
\eqref{Lax}--\eqref{LF} whose curvature is zero when the superstring equations
 of motion are satisfied, at least to quadratic order in fermions.

  For the integrability of the system also the
 inverse statement should be true, namely, that the zero--curvature
 condition should lead to  the superstring equations of motion. In Appendix
  D we will show that this is indeed the case. Let us also note that when
  restricted to the $AdS_2\times S^2$ supercoset sector, the Lax connection
   \eqref{Lax} differs from the one that is usually constructed in the case of the
   supercoset sigma--models \cite{Bena:2003wd,Arutyunov:2008if,
   Stefanski:2008ik,Babichenko:2009dk} by a $PSU(1,1|2)$ gauge
   transformation whose parameters depend on the supercoset
   coordinates $x$ and $\vartheta$, and on the spectral parameter.\footnote{In the
    the $AdS_4\times CP^3$ case the explicit form of a similar  gauge transformation
    relating the two Lax connections was  given in \cite{Sorokin:2010wn}.}

 Let us finish this subsection by  mentioning another similar type IIA  background which is also
 T-dual to   the
type IIB $AdS_2\times S^2\times T^6$  background with the RR  5-form flux
discussed below.
Here $AdS_2\times S^2\times T^6$ is supported by the following 4-form flux
\begin{eqnarray}\label{F4}
F_4&=&
-\frac{e^{-\phi}}{2R}e^be^a\varepsilon_{ab}\I(\Omega_2)
-\frac{e^{-\phi}}{2R}e^{\hat b}e^{\hat a}\varepsilon_{\hat a\hat b}\R(\Omega_2)\,,
\end{eqnarray}
where $\Omega_2$ is the holomorphic 2-form on $T^4\subset T^6$.
This solution describes the near-horizon geometry of the $D2D2D4D4$ brane intersection.
The RR couplings in the GS action for this background are given by
\begin{equation}
\slashed F=
-\frac{4}{R}\gamma\gamma^5\Gamma_{(2)}\mathcal P_8\ .
\end{equation}
with
\begin{equation}
\mathcal P_8=\frac{1}{8}(2+J)\,,\qquad
J=-2\gamma'^5-iJ_{a'b'}\gamma^{a'b'}\gamma^5\,,\qquad J^2=12+4J\,,
\end{equation}
where now  $a',b'$ are  $T^4$ indices, $\gamma'^5$ is the product of the $T^4$ gamma matrices and
$\Gamma_{(2)}=\frac{i}{2}\R(\Gamma^{\mathcal I})\R(\Gamma^{\mathcal J})\varepsilon_{\mathcal{IJ}}$
($\mathcal I,\mathcal J=1,2$ are holomorphic indices),
$\Gamma_{(2)}^2=1$.
This type IIA solution can also be obtained from the $D=11$ supergravity solution
$AdS_2\times S^2\times T^7$ supported by  the same $F_4$-flux by dimensionally
reducing it on an $S^1\subset T^7$. The analysis
analogous to the one carried out above shows that the
GS superstring is integrable also in this  background,
 at least up to the second order in fermions.

\subsection{Type IIB $AdS_2\times S^2\times T^6$ with $F_5$ flux}\label{IIB}

Let us now consider the $AdS_2\times S^2\times T^6$ solution of type IIB
supergravity with RR 5-form flux
\begin{eqnarray}\label{F5}
F_5&=&\frac{e^{-\phi}}{2R}e^be^a\,\varepsilon_{ab}\,\R(\Omega_3)+\mbox{Hodge dual}
\end{eqnarray}
where $\Omega_3=dz^1dz^2dz^3$ is the holomorphic 3-form on $T^6$, $R$ is the
radius of $AdS_2$ and $S^2$ and (as above) $a,b=0,1$,\  $\hat a,\hat b=2,3$ and $a',
b'=4,\ldots,9$ are the
$AdS_2$, $S^2$ and $T^6$ indices respectively.
This solution also preserves eight supersymmetries.\footnote{
Note that as in the type IIA case discussed above,   if we spoil the holomorphic structure
 of $F_5$ by changing signs
 of some of its components, we will get a non--supersymmetric solution. The GS
  string on such a background will, probably, not be integrable.}

Using eq. \eqref{F5} and the definition (\ref{eq:slashedF})
of $\slashed F$ in the IIB case  we find
\begin{equation}\label{IIBB}
\slashed F=\frac{4}{R}\mathcal P_8\sigma^2\gamma\Gamma_{(3)}\frac{1-\Gamma_{11}}{2}\,,
\end{equation}
where $ \mathcal P_8$ is as in \rf{pii}   and
\begin{equation}\label{G(3)}
\Gamma_{(3)}=\frac{i}{3!}\R(\Gamma^{\mathcal I})\R(\Gamma^{\mathcal J})\R(\Gamma^{\mathcal K})\varepsilon_{\mathcal{IJK}}
\end{equation}
and ${\mathcal I},{\mathcal J},\ldots=1,2,3$ refer to holomorphic coordinates on $T^6$ so that
$\Gamma_{(3)}^2=1$ and $[\Gamma_{(3)},\mathcal P_8]=0$.

Let us now  write down
 the (enlarged) \psu algebra \eqref{eq:PSUQ} in a form suitable for this case.
To this end we pass to a different
realization of the gamma--matrices by taking
\begin{eqnarray}\label{redef}
\Gamma^a&\rightarrow&i\Gamma^a\Gamma_{(3)},\nonumber\\
\Gamma^{\hat a}&\rightarrow&i\Gamma^{\hat a}\Gamma_{(3)},\nonumber\\
\R(\Gamma^{\mathcal I})&\rightarrow&\R(\Gamma^{\mathcal I}),\nonumber\\
\I(\Gamma^{\mathcal I})&\rightarrow&i\I(\Gamma^{\mathcal I})\Gamma_{(3)},\nonumber\\
\gamma^7 &\rightarrow & i\gamma^7\,\Gamma_{(3)},
\end{eqnarray}
together with $\mathcal C\rightarrow i\mathcal C\Gamma_{(3)}$.
 This preserves the Clifford algebra, symmetry properties of the
  gamma matrices as well as the form of the projector $\mathcal P_8$,
   though this time in a less trivial way. The fact that the original
   gamma-matrices give $\mathcal P_8\gamma^7\mathcal P_8=i\varepsilon$ means
   that after the redefinition \eqref{redef}, we have $\mathcal P_8\gamma^7
   \mathcal P_8=-\varepsilon\Gamma_{(3)}=-i\sigma^2\Gamma_{(3)}$. After
   replacing in this way $\mathcal P_8\gamma^7\mathcal P_8$ in the (redefined)
   superalgebra \eqref{eq:PSUQ}, we identify the $SO(2)$ index on $Q$ on which
   $\varepsilon=i\sigma^2$ acts with the $SO(2)$ index labeling the Majorana--Weyl
    spinors in type IIB theory. The (enlarged) \psu algebra commutators involving $Q$
    (see Appendix B) then become

\begin{eqnarray}\label{IIBa1}
[P_A,Q]=-\frac{1}{2R}Q\gamma\sigma^2\Gamma_{(3)}\Gamma_A\mathcal
P_8\,,\qquad[M_{\underline{ab}},Q]=-\frac{1}{2}Q\Gamma_{\underline{ab}}\mathcal P_8\,,\qquad[T,Q]=\frac{i}{2}\sigma^2Q\,,
\end{eqnarray}
and
\begin{eqnarray}\label{IIBa2}
\{Q,Q\}=
2i(\mathcal P_8\Gamma^A\mathcal P_8)P_A
-\frac{iR}{2}(\mathcal P_8\Gamma^{AB}\gamma\Gamma_{(3)}\sigma^2\mathcal
P_8)R_{AB}{}^{CD}M_{CD}\,.
\end{eqnarray}
This is the form of the algebra we need in this case since it involves
$\slashed F\propto\mathcal P_8\sigma^2\gamma\Gamma_{(3)}$.

Now the relations between Killing vectors and Killing spinors and the generators
of the algebra are
\begin{eqnarray}
K_A=kP_Ak^{-1},\qquad i\mathcal C\gamma\Gamma_{(3)}\sigma^2\Xi=kQk^{-1}\,,\\
\nabla_AK_B=-\frac{1}{2}R_{AB}{}^{CD}kM_{CD}k^{-1}\,.
\end{eqnarray}
We get
\begin{eqnarray}
[K_A,\Xi]=\frac{1}{2R}\Xi\Gamma_A\sigma^2\gamma\Gamma_{(3)}\mathcal P_8\mathcal C
\end{eqnarray}
and
\begin{eqnarray}
\{\Xi,\Xi\}=
2i(\mathcal P_8\gamma\Gamma^A\gamma\mathcal P_8\mathcal C)\,K_A
-iR(\mathcal P_8\Gamma^{AB}\sigma^2\gamma\Gamma_{(3)}\mathcal
P_8\mathcal C)\,\nabla_AK_B\,.
\end{eqnarray}
Using this algebra together with \eqref{j-ej} and \eqref{susy} we find
\begin{equation}
J_{susy}^2=-\Big[\nabla J^{AB}+e^A(J^B-e^B)\Big]\,\nabla_AK_B
\end{equation}
and, using \eqref{eq:dJSUSY} and \eqref{BC},
\begin{equation}
J_{susy}J_{\mathcal B}+J_{\mathcal B}J_{susy}=-\frac{1}{2}dJ_{susy}\,,
\end{equation}
which is precisely what we need
for the Lax connection \eqref{Lax} to have zero curvature.
This proves the integrability of the GS superstring also in this
background, at least  up to the quadratic order in fermions.

\section{Consistent truncation of the $D=10$ superstring to the\\
  $PSU(1,1|2)/SO(1,1)\times U(1)$ supercoset sigma--model}
\setcounter{equation}0

Let us now show that the GS actions
describing the type IIA  and type IIB  superstrings in the
above  $AdS_2\times S^2\times T^6$ backgrounds
preserving eight supersymmetries can be consistently truncated, at the classical level,
 to the $PSU(1,1|2)/SO(1,1)\times U(1)$ supercoset GS sigma--model with
 the  $AdS_2\times S^2$ bosonic part and  eight fermionic modes.

\subsection{Quadratic fermionic  part of the  superstring  Lagrangian}

The string bosonic modes $x^{\underline m}$
 along $AdS_2\times S^2$ and the eight fermionic modes
$\vartheta=\mathcal P_8\Theta$ together  parameterize the $PSU(1,1|2)/SO(1,1)\times U(1)$ supercoset.
We would like to separate their contribution in the GS action from the coordinates $y^{a'}$ along
$T^6$ and the remaining
24 fermionic modes $\upsilon$. Denoting
\begin{equation}
\upsilon=(1-\mathcal P_8)\Theta\ , \ \ \ \ \ \ \ \ \ \ \ \ \vartheta=\mathcal P_8\Theta \ ,
\end{equation}
and using the properties \eqref{PG} and \eqref{PTP} of the projector $\slashed F\sim \mathcal P_8$
we find for the quadratic
part of the fermion Lagrangian  in  the type IIA
background from sec.~\ref{F2F4}
($h_{ij}$ is the auxiliary worldsheet metric):
\begin{eqnarray}\label{fermLIIA}
 \mathcal{L}_{\rm ferm}
&=&i\vartheta\left(\sqrt{-h}h^{ij}-\varepsilon ^{ij}{\Gamma_{11} }\right)e_i{}^{\underline a}\Gamma_{\underline a}
 \,
 \left(\nabla_j+\frac{1}{2R}\mathcal{P}_8\gamma \Gamma _{11}\Gamma_{\underline a} e_j{}^{\underline a}\right)\,\vartheta\nonumber\\
 &&+i\vartheta \left(\sqrt{-h}h^{ij}-\varepsilon ^{ij}\Gamma_{11}\right)\Gamma_{a'}\nabla_j\upsilon \,\partial_iy^{a'}
 +i\upsilon \left(\sqrt{-h}h^{ij}-\varepsilon ^{ij}\Gamma_{11}\right)\Gamma_{a'}\nabla_j\vartheta \,\partial_iy^{a'}\nonumber\\
&&{}+\frac{i}{R}\vartheta\left(\sqrt{-h}h^{ij}-\varepsilon ^{ij}{\Gamma_{11} }\right)e_i{}^{\underline a}\Gamma_{\underline a}\,\mathcal{P}_8\gamma \Gamma _{11}\Gamma_{a'}
\upsilon\,\partial_jy^{a'}\\
&&{}+i\upsilon\left( \sqrt{-h}h^{ij}-\varepsilon ^{ij}{\Gamma_{11} }\right)\slashed e_i
 \nabla_j\upsilon
 +\frac{i}{2R}\,\upsilon\left(
 \sqrt{-h}h^{ij}-\varepsilon ^{ij}{\Gamma_{11} }
 \right)\Gamma_{a'}\mathcal{P}_8\gamma \Gamma _{11}\Gamma_{b'}\upsilon\,\partial_i y^{a'}\partial_j y^{b'},\nonumber
\end{eqnarray}
where
\begin{equation}\label{defsforGSIIA}
\slashed e_i=\Gamma _{\underline{a}}e_i{}^{\underline{a}}(x)+\Gamma_{a'}\partial_i y^{a'},\qquad
\nabla_j=\partial _j+\frac{1}{4}\,\Gamma _{\underline{ab}}\,\omega_j{}^{\underline{ab}}(x)\,,
\end{equation}
and $\omega _i{}^{\underline{ab}}(x)$ and $e_i{}^{\underline{a}}(x)$ are
 the worldsheet pull-backs of the spin connection and the local frame
  in $AdS_2\times S^2$.

{We immediately see that the coset and non--coset directions do not  decouple
from each other.
The couplings between the respective worldsheet fields (or their derivatives)  have the following schematic
form: $\upsilon y\vartheta$, $\upsilon x\upsilon$, $\vartheta yy \vartheta$,
$\vartheta yx\upsilon$ etc.
 The non--coset fields, however, never appear linearly in the Lagrangian, and can thus be set to zero consistently with the equations of motion, which always admit the trivial solution $y^{a'}=0$, $\upsilon=0$. This crucially depends on the absence of linear couplings
where only a single $y$ or $\upsilon$  field would
appear together with $x$ and $\vartheta$.\footnote{The absence of such couplings and the structure of all other terms in the action is related to its invariance under the external $U(1)$--automorphism of the \psu algebra generated by the K\"ahler form $J_{a'b'}$ on $T^6$ (see Appendix B). With respect to this $U(1)$ the $x$ coordinates are neutral, the three complex coordinates $y$ of $T^6$ have charge 1, the complex counterparts of $\vartheta$ have charge $3/2$ and the $\upsilon$'s have charge $-1/2$. The ratio $-3$ of the charges of $\vartheta$ and $\upsilon$ is explained by the fact that they are eigenfunctions of the matrix $J=-iJ_{a'b'}\,\gamma^{a'b'}\,\gamma^7 $, appearing in the projector $\mathcal P_8$, whose eigenvalues are $6$ with degeneracy 8 (associated with $\vartheta$) and $-2$ with degeneracy 24 (associated with $\upsilon$). One can see that these values of the charges forbid any term in the action with a single $y$ or $\upsilon$, provided that only neutral terms appear.}

Thus, in  the case under consideration
one can  consistently
reduce the classical theory to the
 $AdS_2\times S^2$
sector  coupled to 8 fermions $\vartheta$. As we shall see below,
this  reduced theory   is nothing but the
$PSU(1,1|2)/SO(1,1)\times U(1)$ supercoset sigma model.
 Due to
  the presence of the 4--component kappa--symmetry,
   the number of physical  fermionic  degrees of
   freedom in this model is equal to two, i.e. is the same as the number
   of physical bosonic degrees of
   freedom.\footnote{Let us also note
that when the string moves in
$T^6$ (i.e. the  coordinates $y$ are not constant) it is possible to
gauge-fix the eight fermions $\vartheta$ to zero using half of the
kappa--symmetries. In such a gauge one is left with 24 broken supersymmetry
fermions $\upsilon$ and eight  transverse worldsheet bosonic modes $x$ and
$y$. It is not possible, however, to further consistently truncate the theory
by freezing some of the transverse excitations $x$ and/or $y$ because of the presence of  cubic terms of the form $\upsilon x\upsilon$ and $\upsilon
y\upsilon$ in
the action. Notice that in this case we still have eight remaining
kappa--symmetries  so that the number of physical (on-shell)  bosonic and fermionic degrees of
freedom is  of course 8+8.}

Let us now keep only the $x$ and $\vartheta$ fields in the GS Lagrangian:
\begin{equation}\label{1st}
\mathcal{L}^{(2)}_{\rm GS}
=i\vartheta\left(\sqrt{-h}h^{ij}-\varepsilon ^{ij}{\Gamma_{11} }\right)e_i{}^{\underline a}\Gamma_{\underline a}
 \,
 \left(\nabla_j+\frac{1}{2R}\mathcal{P}_8\gamma \Gamma _{11}\Gamma_{\underline a} e_j{}^{\underline a}\right)\,\vartheta.
\end{equation}
We are going to identify
the Dirac matrices that appear in this Lagrangian with the
structure constants of the $\mathfrak{psu}(1,1|2)$ superalgebra
in a suitable realization
 (see Section \ref{F2F4} and Appendix B).

Since $AdS_2\times S^2$ is a coset $SO(1,2)\times SO(3)/SO(1,1)\times SO(2)$ its
 spin connection and the vielbein appear in the expansion of the left--invariant
  current in the algebra generators:
\begin{equation}\label{decompos}
 g^{-1}\partial _ig=\frac{1}{2}\,
 \omega_i{}^{\underline{ab}}
  M_{\underline{a}\underline{b}}+e_i{}^{\underline{a}}P_{\underline{a}}
 \equiv A_i+P_i,
\end{equation}
where $g\in SO(1,2)\times SO(3)$ is a coset representative which defines an embedding of $AdS_2\times S^2$ into $ SO(1,2)\times SO(3)$, and $\left\{P_{\underline{a}},M_{\underline{a}\underline{b}}\right\}$ are the generators
 of $\mathfrak{so}(1,2)\oplus\mathfrak{so}(3)$, defined in Appendix~B.
The gamma--matrices, which contract with
the spin connection and the vielbein in (\ref{defsforGSIIA})
 form a representation of $\mathfrak{so}(1,2)\oplus\mathfrak{so}(3)$:
\begin{equation}
 \left(P_{\underline{a}},M_{\underline{a}\underline{b}}\right)\rightarrow
 \left(\frac{1}{2R}\,\gamma \Gamma _{11}\Gamma _{\underline{a}},
 -\frac{1}{2}\,\Gamma _{\underline{a}\underline{b}}
 \right).
\end{equation}
The same representation appears
 in the $D=10$ form (\ref{repPS112}) of the $\mathfrak{psu}(1,1|2)$  algebra. If we define the $\mathfrak{psu}(1,1|2)$--valued fields
\begin{equation}\label{TQ}
 \Psi=\vartheta Q,
\end{equation}
(where $Q$ are the supersymmetry generators) the spin connection and the local frame
will act
 on $\Psi$ as the Lie--algebra commutators:
\begin{equation}
 \frac{1}{2R}\,\gamma \Gamma _{11}\Gamma_{\underline a} e_i{}^{\underline a}\rightarrow
 \left[P_i,\ \cdot\right],\qquad
 -\frac{1}{4}\,\Gamma _{\underline{ab}}\,
  \omega_i{}^{\underline{ab}}\rightarrow
  \left[A_i,\ \cdot\right]\,.
\end{equation}

We also introduce a parity transformation $\Omega$ of the fermion field:
\begin{equation}\label{Ops}
\qquad \Omega\Psi=-i\vartheta\Gamma_{11}Q\,.
\end{equation}
Later we will identify it with the
 $\mathbb Z_4$--automorphism of the $\frak{psu}(1,1|2)$ superalgebra.

The Lagrangian (\ref{1st}) then takes the form
\begin{equation}\label{fermLIIA1}
 \mathcal{L}^{(2)}_{\rm GS}
 =\frac{1}{2}\mathop{\mathrm{Str}}
 \Psi\left[P_i,\left(\sqrt{-h}h^{ij}+i\varepsilon ^{ij}\Omega \right)
 \mathcal{D}_j\Psi
 \right],
\end{equation}
where
\begin{equation}
  \mathcal{D}_j\vartheta=\nabla_j\vartheta+[P_j,\vartheta], \qquad
  \nabla_j\vartheta=\partial _j\vartheta+[A_j,\vartheta],
\end{equation}
and $\mathop{\mathrm{Str}}(\cdot\,,\cdot)$ is the
 invariant bilinear form on $\mathfrak{psu}(1,1|2)$,
 which coincides with
   $\mathcal{C}\gamma \Gamma _{11}$
  on the 8--dimensional spinor space $Q=\mathcal P_8Q$.
   More precisely,
 \begin{equation}
  \mathrm{Str}\,(Q_\alpha Q_\beta)=
    -4iR(\mathcal C\gamma\Gamma_{11}\mathcal P_8)_{\alpha\beta}
\end{equation}
   in the normalization in which  $\mathrm{Str}\,(P_{\underline a}P_{\underline b})=
    \eta_{\underline{ab}}$.

We can now compare \eqref{fermLIIA1} with
the Lagrangian
of the $PSU(1,1|2)/SO(1,1)\times U(1)$ coset sigma--model expanded to the second order in fermions.

\subsection{Lagrangian of the supercoset model}

The $PSU(1,1|2)/SO(1,1)\times U(1)$ coset is a semi--symmetric superspace
\cite{Serganova},
which means that its supergeometry is invariant under the action of a $\mathbbm{Z}_4$--symmetry. This symmetry is inherited from the
$\mathbbm{Z}_4$--automorphism of $\mathfrak{psu}(1,1|2)$ that acts
on the generators, in the
$D=10$
 notation of (\ref{repPS112}) and Appendix B, as follows
\begin{equation}\label{Z4auto}
 \Omega (M_{\underline{a}\underline{b}})=M_{\underline{a}\underline{b}},
 \qquad
 \Omega (P_{\underline{a}})=-P_{\underline{a}},
 \qquad
 \Omega (Q)=-iQ\Gamma _{11}.
\end{equation}
The action of $\Omega$ preserves the commutation relations of the superalgebra
and thus defines a $\mathbbm{Z}_4$--decomposition of $\mathfrak{psu}(1,1|2)$:
\begin{equation}
 \mathfrak{psu}(1,1|2)=\mathfrak{h}_0\oplus\mathfrak{h}_1\oplus\mathfrak{h}_2\oplus\mathfrak{h}_3,\qquad \Omega (\mathfrak{h}_p)=i^p\mathfrak{h}_p,
 \qquad [\mathfrak{h}_p,\mathfrak{h}_q\}\subset\mathfrak{h}_{(p+q)\!\!\!\!\mod 4}.
\end{equation}
The invariant subspace of $\Omega $, $\mathfrak{h}_0=\mathfrak{o}(1,1)\oplus\mathfrak{u}(1)$,
 is the denominator subalgebra of the supercoset.
 {The action of the $\mathbbm{Z}_4$ automorphism on the supercharges coincides with the 10d parity (cf. eq.\eqref{Ops}).

The string embedding in $PSU(1,1|2)/SO(1,1)\times U(1)$  is parameterized by a
coset representative $G(\xi)\in
PSU(1,1|2)$, where $\xi$ are the worldsheet coordinates. The coset representative is defined up to local $SO(1,1)\times U(1)$ transformations: $G(\xi)
\rightarrow G(\xi) h(\xi)$. The global $PSU(1,1|2)$ symmetry acts from the left: $G(\xi)\rightarrow gG(\xi)$. The Lagrangian
is constructed from the left-invariant current
\begin{equation}\label{coset_current}
 J_j=G^{-1}\partial _jG,
\end{equation}
with the help of the $\mathbbm{Z}_4$--decomposition:
\begin{equation}\label{coset_decomp}
 J_{j\,p}=\frac{1}{4}\sum_{l=0}^{3}i^{-lp}\Omega
 ^l(J_j)\in\mathfrak{h}_p\ .
\end{equation}

The $\mathfrak{h}_0$--projection is the $SO(1,1)\times U(1)$ gauge field
 (or, equivalently, the $AdS_2\times S^2$ spin connection), since it transforms
  as $J_{j\,0}\rightarrow h^{-1}J_{j\,0}h+h^{-1}\partial _jh$. The other components of the current transform in the adjoint of $SO(1,1)\times U(1)$: $J_{j\,1,2,3}\rightarrow h^{-1}J_{j\,1,2,3}h$. The Lagrangian of the supercoset sigma-model is
\begin{equation}\label{cosetL}
 \mathcal{L}_{\rm coset}=\frac{1}{2}\mathop{\mathrm{Str}}\left(\sqrt{-h}h^{ij}J_{i\,2}
 J_{j\,2}+\varepsilon ^{ij}J_{i\,1}J_{j\,3}\right).
\end{equation}

In order to compare the sigma--model with the GS string Lagrangian \eqref{1st},
we should expand the sigma--model Lagrangian to the second order in fermions.
 This can be understood as a particular case of the background field expansion,
 the general form of which is given in Appendix~B of \cite{Babichenko:2009dk}.
 The starting point is the following form of the coset representative
\begin{equation}
 G=g\,{\rm e}\,^{\mathbbm{X}},
\end{equation}
 where $g(x)\in SO(1,2)\times SO(3)$
 is a bosonic element of the supergroup that describes a classical string solution
 in $AdS_2\times S^2$.
 The field $\mathbbm{X}\in \mathfrak{h}_1
 \oplus\mathfrak{h}_2\oplus\mathfrak{h}_3$ describes string fluctuations around this
 solution. The coset gauge is fixed by requiring that $\mathbbm{X}$ has zero $\frak{h}_0$ projection.

 The sigma--model Lagrangian, expanded to the second order
  in fluctuations, takes the following form (upon dropping a total derivative coming from the WZ term of \eqref{cosetL}):
\begin{eqnarray}\label{fluct}
 \mathcal{L}_{\rm fluct}&=&\frac{1}{2}\mathop{\mathrm{Str}}\left\{
 \sqrt{-h}h^{ij}\left(\nabla_i\mathbbm{X}_2\nabla_j\mathbbm{X}_2-[P_i,\mathbbm{X}_2][P_j,\mathbbm{X}_2]\right)
 \right.
\nonumber \\
&&\left.
 +\left(\sqrt{-h}h^{ij}+\varepsilon ^{ij}\right)
 \mathbbm{X}_1\left[P_i,\nabla_j\mathbbm{X}_1+[P_j,\mathbbm{X}_3]\right]
  \right.
\nonumber \\
&&\left.
 +\left(\sqrt{-h}h^{ij}-\varepsilon ^{ij}\right)
 \mathbbm{X}_3\left[P_i,\nabla_j\mathbbm{X}_3+[P_j,\mathbbm{X}_1]\right]
 \right\},
\end{eqnarray}
where $A_j$ and $P_j$ are grade $0$ and grade $2$ components of
 the $\mathbbm{Z}_4$--decomposition of the background Cartan form as in
  (\ref{decompos}), and the covariant
 derivative contains the background gauge connection $\nabla_j=\partial _j+[A_j,\cdot]$.
The fermion part of the fluctuation Lagrangian can be more compactly written in terms of
\begin{equation}
\Psi =\mathbbm{X}_1+\mathbbm{X}_3,\qquad \mathbbm{X}_{1,3}=\frac{1\pm i\Omega }{2}\,\Psi .
\end{equation}
Namely,
\begin{equation}\label{2coset}
 \mathcal{L}^{(2)}_{\rm coset}=
 \frac{1}{2}\mathop{\mathrm{Str}}
 \Psi \left[P_i,\left(\sqrt{-h}h^{ij}+i\varepsilon ^{ij}\Omega \right)
 \left(\nabla_j\Psi +[P_j,\Psi ]\right)\right]
\end{equation}
which coincides with \eqref{fermLIIA1}.

For the type IIB background (\ref{F5}), the part of the GS action that contains
the $PSU(1,1|2)$ fields becomes
\begin{equation}\label{GreenSchwarzIIB}
 \mathcal{L}^{(2)}_{\rm coset}=i\vartheta\left(
 \sqrt{-h}h^{ij}-\varepsilon ^{ij}\sigma ^3
 \right)e_i{}^{\underline a}\Gamma_{\underline a}\left(
 \nabla_j-\frac{1}{2R}\,\mathcal{P}_8\sigma ^2\gamma \Gamma _{(3)}e_j{}^{\underline b}\Gamma_{\underline b}
 \right)\vartheta
\end{equation}
with the $\mathfrak{so}(1,2)\times \mathfrak{so}(3)$ generators being
\begin{equation}
 \left(P_{\underline{a}},M_{\underline{a}\underline{b}}\right)\rightarrow
 \left(\frac{1}{2R}\,\sigma ^2\gamma \Gamma _{(3)}\Gamma _{\underline{a}},
 -\frac{1}{2}\,\Gamma _{\underline{a}\underline{b}}\right).
\end{equation}
This coincides with the representation of the bosonic generators
 of $\mathfrak{psu}(1,1|2)$ in the type IIB
 realization of the superalgebra (\ref{IIBa1})--\eqref{IIBa2}. Finally,
 the $\mathbbm{Z}_4$--automorphism acts on the supercharges as
\begin{equation}
 \Omega (Q)=iQ\sigma ^3,
\end{equation}
 and thus (\ref{GreenSchwarzIIB}) is equivalent to the
 coset action (\ref{2coset}) for the fields that satisfy $\mathcal{P}_8\Theta =\Theta$.

\subsection{Supercoset truncation beyond the quadratic approximation}

Let us now consider how one may, in principle, reconstruct the full $AdS_2\times S^2\times T^6$ supergeometry by solving the supergravity constraints order by order in non--supersymmetry fermions $\upsilon$, with the $PSU(1,1|2)/SO(1,1)\times U(1)$ geometry taken as the initial condition.
This will also demonstrate that the supercoset sigma--model action is a consistent truncation of the full
non--linear type II superstring action.

Let us now demonstrate that the $PSU(1,1|2)/SO(1,1)\times U(1)$ supercoset action
is a consistent truncation of the full
non--linear type II superstring action.

The GS superstring action in a generic type II D=10 superbackground is \cite{Grisaru:1985fv}
\begin{equation}\label{cordaA}
S = -\frac{1}{4\pi\alpha'}\,\int *E^AE^B
\eta_{AB} -\frac{1}{2\pi\alpha'}\,\int  B_2\,,
\end{equation}
where
$E^A(X,\Theta)=dX^M\,E_M{}^{A}+d\Theta^{\underline\mu}E_{\underline\mu}{}^A$
are the worldsheet
pullbacks of the vector supervielbein and the  2-form field
$B_2(X,\Theta)$ can be computed from its field strength $H_3=dB_2$ as
\begin{equation}\label{B2}
B_2=b_2(X)+\int_0^1\,dt\,i_\Theta H_3(X,t\Theta)\,.
\end{equation}
The vielbeins satisfy the type II supergravity torsion constraint\footnote{The
superspace constraints used here differ from those used in \cite{Gomis:2008jt,Grassi:2009yj} by a shift in the fermionic supervielbein, $E=\mathcal E+\frac{1}{2}E^A\,\Gamma_A\lambda$, where $\lambda$ is the dilatino superfield.}
\begin{eqnarray}\label{Ta=IIB}
T^A=DE^A=dE^A-E^B\,\Omega_B{}^A=-iE\Gamma^AE\,,
\end{eqnarray}
where $\Omega_B{}^A(X,\Theta)$ is the spin connection,
$E^{\underline\alpha}$ are the 32--component spinor supervielbeins of the opposite
(type IIA) or the same (type IIB) chirality. In the linear approximation
$E^{\underline\alpha}=\mathcal D\Theta^{\underline \alpha}$, where $\mathcal D$ is a covariant
derivative similar to \eqref{E}.
$H_3$ satisfies the following constraint
\begin{eqnarray}\label{H3=IIBstring}
H_3=dB_2=
iE^A\,E\Gamma_A\hat\Gamma E
+\frac{1}{3!}E^CE^BE^A\,H_{ABC}\,.
\end{eqnarray}
Let us now concentrate on the $AdS_2\times S^2\times T^6$ case. To prove that the supercoset model is a consistent truncation we should
know the dependence
of the supervielbeins on the 8 supercoset fermions $\vartheta$ (which in general
extends up  to the
eighth order) and their
dependence on the extra 24 fermions $\upsilon$ and the derivatives of the
$T^6$ coordinates $y$ to
linear order. This dependence can be determined by a recursion procedure
similar to the one used, e.g., in
\cite{deWit:1998tk,Claus:1998fh} (and references there)
  assuming  the $PSU(1,1|2)/SO(1,1)\times U(1)$ supergeometry as the initial
condition.

One can always impose a gauge condition for the target superspace
diffeomorphisms such that
\begin{equation}\label{gf}
\upsilon^{\underline\alpha}\,E_{\underline\alpha}{}^A\equiv i_\upsilon E^A=0,
\qquad \upsilon^{\underline\alpha}\,\Omega_{\underline\alpha}{}^{AB}
\equiv i_\upsilon \Omega^{AB}=0\ ,
\end{equation}
where  $\upsilon$ are again  the 24 non--supercoset fermions
$\upsilon=(1-\mathcal P_8)\Theta$ associated to the broken supersymmetries.

The torsion constraint \eqref{Ta=IIB}, the $PSU(1,1|2)$ superisometry of the
background and the
$\Gamma$-matrix relations \eqref{PG} and \eqref{PTP} then imply that the
vector supervielbeins have the
following structure. Along the coset direction $AdS_2\times S^2$ (which is
labeled by $\underline
c$) the supervielbein is
\begin{equation}\label{C}
E^{\underline c}=e^{\underline c}(x,\vartheta)+i\upsilon\Gamma^{\underline c}
\mathcal D\upsilon+\mathcal O(\upsilon^4)\ ,
\end{equation}
where $e^{\underline c}(x,\vartheta)$ is  the complete expression for the
supervielbein of the $PSU(1,1|2)/SO(1,1)\times U(1)$ supercoset
(which can be computed from the Cartan form). Note that, by construction, the components
of this  supervielbein satisfy a
condition similar to \eqref{gf}
\begin{equation}\label{gfc}
\vartheta^\mu\,e_{\mu}{}^{\underline c}(x,\vartheta)\equiv
i_\vartheta\,e^{\underline c}=0\,.
\end{equation}
In the supervielbein along the coset directions there are no terms linear in
the extra 24 fermions
$\upsilon$. The absence of such terms is guaranteed by the torsion constraint
\eqref{Ta=IIB} and the gamma-matrix identities
\eqref{PG}--\eqref{PTP}.

Along the $T^6$ directions the supervielbein has the following form
\begin{equation}\label{T}
E^{a'}=dy^{a'}+2i\upsilon\Gamma^{a'}e(x,\vartheta)+\mathcal O(\upsilon^2)\,,
\end{equation}
where $e^{\underline\alpha}(x,\vartheta)=(\mathcal P_8e)^{\underline\alpha}$ is
the spinorial supervielbein  of
the supercoset (contained in the Cartan form).
Because of the $\mathcal P_8$ projector the spinorial index $\underline \alpha$ here takes 8 values.

The 8 fermionic  supervielbeins
 of the complete target superspace (associated with
the supercoset) start from the corresponding supercoset
fermionic Cartan form $e^{\underline\alpha}(x,\vartheta)$ and may contain terms quadratic
in the fields $\upsilon$ and
$dy$
\begin{equation}\label{fer}
(\mathcal P_8E)^{\underline \alpha}=e^{\underline\alpha}(x,\vartheta)+(\mathcal P_8 dy^{a'}\Gamma_{a'}\,\upsilon)^{\underline\alpha} + \mathcal O(\upsilon^2)\,.
\end{equation}
The extra 24 fermionic supervielbeins are:
\begin{equation}
(1-\mathcal P_8)E=(1-\mathcal P_8)\mathcal D\upsilon+\mathcal O(\upsilon^3)\,.
\end{equation}
To get the explicit form of the spinorial supervielbeins to all orders in
$\upsilon$ one should use the
explicit form of the type II spinorial torsion
$DE^{\underline\alpha}=T^{\underline\alpha}$ in the $AdS_2\times
S^2\times T^6$ backgrounds and Bianchi identities for the relevant superforms.
 The reconstruction of  this dependence is technically rather
 involved and so far has not been carried out.

Finally, the  term which might spoil the consistent truncation to
the supercoset model could
be the contribution to $B_2$ \eqref{B2} which comes from the second term in
\eqref{H3=IIBstring} with all the indices along the coset directions, \emph{i.e.}
$\vartheta^\mu e_\mu{}^{\underline a}e^{\underline b}e^{\underline c}
H_{\underline{abc}}$. Note that in the supercoset model the three-form field
 strength $H_{\underline{abc}}$ is zero,
while in the complete
$D=10$ superbackground $H_{\underline{abc}}$ may, in principle, depend
linearly on $\upsilon$. However,  this potentially
dangerous term in the string action is zero because of eq. \eqref{gfc}.
This is in accordance with the $U(1)$--automorphism invariance discussed in footnote 15.
Thus,
with the above form of the supervielbeins one can use the same reasoning as in
the quadratic approximation to show that the GS action can be consistently
truncated to the complete non--linear supercoset model.


\section{BMN limits of the $AdS_2\times S^2\times T^6$ superstring
}\label{BMN}
\setcounter{equation}0

By analogy with the BMN expansion  of the $AdS_5\times S^5$ superstring (in which, in particular,
 the unbroken $PSU(2|2)\times PSU(2|2)$ symmetry plays an  important role, see,
 e.g., \cite{Arutyunov:2009ga})
 it is interesting to study
  the BMN limits of the $AdS_2\times S^2\times T^6$ superstring,
   having in mind their possible role in its  exact solution based on integrability.

  In general, to  describe the spectrum of  GS strings one may, as in flat space,
    fix  a kind of  light-cone gauge  concentrating on
    a subsector of string states carrying momentum
   in a particular compact direction. In the point-like  limit the string is then
    moving along a
   particular $D=10$  null geodesic which runs along global time direction in AdS
   and along a    geodesic  in a compact direction. The BMN limit corresponds
   to expanding near this geodesic.  In the $AdS_2\times S^2\times T^6$ case
   there  are several inequivalent choices for the geodesic, e.g., it may  run (i)  along big circle
   of $S^2$, or  (ii)  along $S^1$ of $S^2$  and $S^1$ of  $T^6$ at the same time
   (with the case  of only $S^1$ of  $T^6$ being a special case  of the latter).
   These lead  to inequivalent BMN limits that we will consider in detail below.

Before getting into a detailed discussion of BMN limits, let  us  mention that
the quadratic supercoset Lagrangian (\ref{fluct}) is very convenient for analyzing the
 symmetries and the spectrum of the string in the first
 BMN limit  when the geodesic runs along $S^2$ in  $AdS_2\times S^2$.
  To this end,
  it is convenient to represent  $\mathfrak{psu}(1,1|2)$ elements by $(2|2)\times (2|2)$
   supermatrices. The BMN background  is represented by \foot{This   BMN limit in
  the $AdS_2\times S^2$ supercoset subsector of the theory was discussed from a
  superalgebra point of view  in \cite{Hatsuda:2003er}.}
   \begin{equation}
   g=\exp(i\tau P_+)\ , \ \ \ \ \ \ \ \ \ \
   P_+=\mathop{\mathrm{diag}}(1,-1|1,-1)\in \mathfrak{h}_2  \ .
   \end{equation}
    Then $A_i=0$ and
     $P_i\sim P_+$. From (\ref{fermLIIA1}) it follows that the components of $\Psi$ (or
     $\vartheta$) that commute with $P_+$ completely drop out of the action  which  is
     a manifestation of the kappa--symmetry. Only those components of the fermion fields
     that have non--trivial commutators with $P_+$ correspond to the physical on--shell
     excitations of the string. Since, the commutant of $P_+$ contains exactly half of
      the supercharges of $\mathfrak{psu}(1,1|2)$, the kappa--symmetry reduces the number
      of the $\vartheta$--fermion fields from 8 to 4 that describe two on--shell fermion states.
The mass--squared matrix for the remaining
coset excitations (both bosonic and fermionic) is then (see (\ref{fluct}))
\begin{equation}
 \mathcal{M}^2=\left(\mathop{\mathrm{ad}}P_+\right)^2  \ .
\end{equation}
We thus get  two bosons plus two fermions of  equal mass.
The commutant of $P_+$  corresponds to the symmetries which are left unbroken by the
background. Here  this  commutant  is $\mathfrak{su}(1|1)\times \mathfrak{su}(1|1)$,
 where $\mathfrak{su}(1|1)$ is  an algebra of two real supercharges whose anti--commutator
 is a bosonic central element.

 As we shall see below, in this BMN limit the complete $D=10$ superstring has an
 additional  six massless bosons and six massless fermions from the
  bosonic fluctuations on $T^6$ and
  non-supersymmetric $\upsilon$--fermions.
  In the second  BMN limit involving $T^6$   we shall find that the masses
  of world-sheet  bosons and fermions will be different, i.e. there will be no
  2d supersymmetry at the quadratic level of the expanded string action.
Moreover, the coset fermions $\vartheta$ will mix with the broken--supersymmetry fermions
$\upsilon$.

\subsection{BMN  limit along   geodesic in $S^2$}\label{5.1}

BMN (or pp-wave)  limits for $AdS_2\times S^2\times M^6$ backgrounds
and
 their supersymmetries were considered, e.g.,   in \cite{Cvetic:2002si}.
   To take the limit,
   let us choose the  $AdS_2\times S^2$ coordinates as
and  the metric of $AdS_2\times S^2\times T^6$ as
\begin{equation}\label{met}
ds^2=R^2(-\cosh^2\rho\,dt^2+d\rho^2+d\theta^2+\sin^2\theta\,d\varphi^2)+dy_{a'}dy^{a'}\,.
\end{equation}
Consider a particle moving with the speed of light in the $\varphi$--direction, set
 $\tilde x^{\pm}=\frac{t\pm\varphi}{2}$ and zoom--in on the geometry
seen by the particle by letting
\begin{equation}
x^+=\frac{\tilde x^+}{\mu}\,,\qquad x^-=\mu R^2\tilde x^-\,,\qquad \rho=\frac{r}{R}\,,
\qquad\theta=\frac{\pi}{2}+\frac{z}{R}\,,\qquad \mu R,\,\frac{R}{r},
\,\frac{R}{z}\rightarrow\infty
\,,
\end{equation}
where $\mu$ is a  mass scale parameter. Taking this limit the metric becomes
\begin{equation}\label{ppm}
ds^2=-4dx^+dx^--\mu^2(r^2+z^2)(dx^+)^2+dr^2+dz^2+dy_{a'}dy^{a'}\,.
\end{equation}
For the type IIA solution  the fluxes \rf{RR} become
\begin{eqnarray}
F_2&=&e^{-\phi}\mu\,dr\,dx^+\\
F_4&=&e^{-\phi}\mu\,dz\,dx^+\,J_2\,,
\end{eqnarray}
while for the type IIB solution we get from \rf{F5}
\begin{equation}
F_5=-e^{-\phi}\mu\,dr\,dx^+\,\mathrm{Re}(\Omega_3)+\mbox{Hodge dual}.
\end{equation}

\subsubsection{Supersymmetries of the pp-wave background}

To see how many supersymmetries these backgrounds preserve
we should analyze the dilatino and gravitino variations.
In our case these take the form
\begin{eqnarray}\label{dilatino}
\Gamma^A\slashed F\Gamma_A\epsilon&=&0\\
(\nabla_A-\frac{1}{8}\slashed F\Gamma_A)\epsilon&=&0\,,\label{gravitino}
\end{eqnarray}
where the second is simply the Killing spinor equation.

For the type IIA pp-wave we get
\begin{equation}
\label{eq:IIAFslash}
\slashed F=-4i\mu\gamma^7\Gamma^z\mathcal P_8\Gamma^+\,.
\end{equation}
This background thus allows 16 Killing spinors which are annihilated by $\Gamma^+=
\frac{1}{2}(\Gamma^0+\Gamma^3)$. It can be shown that these supersymmetries will
always be present in the pp-wave (Penrose)  limit of any background.

 Since $\Gamma^A\slashed F\Gamma_A=
\ldots\Gamma^+(1-\mathcal P_8)$ the dilatino equation also allows for four additional spinors
satisfying $\epsilon=\mathcal P_8\epsilon$ and $\Gamma^+\epsilon\neq0$. To see whether these
 correspond to supersymmetries we need to solve
  also  the gravitino equation \eqref{gravitino} for which
 the integrability condition  reads
\begin{equation}\label{ic}
(R_{AB}{}^{CD}\Gamma_{CD}-\frac{1}{8}\slashed F\Gamma_{[A}\slashed F\Gamma_{B]})\epsilon=0\,.
\end{equation}
To compute the curvature we chose the vielbeins in the form
\begin{equation}
e^+=2dx^+\,\qquad e^-=2(dx^-+\mu^2\frac{r^2+z^2}{4}dx^+), \qquad e^r=dr,\qquad e^z=dz,\qquad e^{a'}=dy^{a'}\,.
\end{equation}
For this choice of the vielbeins the spin connection has the following non--zero components,
defined from the zero--torsion condition $de^A+e^B\omega_B{}^A=0$, \footnote{Note that in all
 the following equations the indices $(+,-)$ refer to the coordinate basis $(dx^+,dx^-)$ and
  not the $(e^+,e^-)$-basis.}
\begin{equation}
\omega^{-r}=-\frac{\mu^2}{2}r\,dx^+\,,\qquad \omega^{-z}=-\frac{\mu^2}{2}z\,dx^+\,.
\end{equation}
Thus, the curvature has the following non--zero components
\begin{equation}
R^{-r}=\frac{\mu^2}{2}dr\,dx^+\,,\qquad R^{-z}=\frac{\mu^2}{2}dz\,dx^+\,.
\end{equation}
The integrability condition \eqref{ic} is thus  trivially satisfied except for the $(+r)$ and
$(+z)$--components. For the $(+r)$--component    \eqref{ic}  gives
\begin{equation}
(2R_{+r}{}^{-r}\Gamma_-\Gamma^r+2\mu^2\Gamma^+\Gamma^r)\mathcal P_8\epsilon=0\,,
\end{equation}
and  the two terms here  cancel each other due to the form of the curvature.
 The $(+z)$-component
vanishes in the same way. This means that the eight supersymmetries of the $AdS_2\times
S^2\times T^6$ background are preserved by the Penrose limit and in addition the system
acquires an extra 12 giving a total of 20 supersymmetries.

In the type IIB case we get
\begin{equation}
\label{eq:IIBFslash}
\slashed F=4\mu\Gamma^r\Gamma_{(3)}\sigma^2\mathcal P_8\Gamma^+\,,
\end{equation}
and again we have at least the 16 Killing spinors which are annihilated by $\Gamma^+$.
In the type IIB case
$\Gamma^A\slashed F\Gamma_A=0$,
and thus there are no terms
in the supersymmetry variation of the dilatino \eqref{dilatino} that depend on $F_5$.
To see whether there are any additional supersymmetries we need to study further the
gravitino equation. Let us first look for the supersymmetries  with $\epsilon=\mathcal
P_8\epsilon$ which were present before the Penrose limit. The non-trivial components of
 the integrability condition \eqref{ic} are again the $(+r)$ and $(+z)$ components. For
 the first of these we get
\begin{equation}
(2R_{+r}{}^{-r}\Gamma_-\Gamma^r+2\mu^2\Gamma^+\Gamma^r)\epsilon=0\,,
\end{equation}
and similarly for the $(+z)$--component. The two terms cancel each other and we again
conclude that the Penrose limit preserves the supersymmetries present in the original
 space, i.e. those with $\epsilon=\mathcal P_8\epsilon$ (only half of which are annihilated
  by $\Gamma^+$).\footnote{For completeness
  we should also check whether other supersymmetries of
  the form $\epsilon=(1-\mathcal P_8)\epsilon$ with $\Gamma^+\epsilon\not=0$ could be present.
 Starting   with  the $(a'+)$-component of the
    integrability condition
$\mathcal P_8\Gamma_{a'}\Gamma^+(1-\mathcal P_8)\epsilon=0\,$
and multiplying it  by $\Gamma^{a'}$ and using  that $\Gamma^{a'}\mathcal
P_8\Gamma_{a'}=2(1-\mathcal P_8)$ (following  from \eqref{P8})  we get
$\Gamma^+(1-\mathcal P_8)\epsilon=0\,  $
which contradicts our assumption, i.e. there are no additional
 supersymmetries.}

We  have thus seen that both the type IIA and type IIB pp--wave backgrounds preserve 20
 supersymmetries: 8  original $AdS_2\times S^2\times T^6$ supersymmetries with
 $\epsilon=\mathcal P_8\epsilon$ and 12  additional ones with $\epsilon=-\Gamma^+\Gamma^-\epsilon$.\footnote{
  Note that the projectors $\mathcal P_8$ (having 8 non--zero eigenvalues) and $\Gamma^+\Gamma^-$
   (having 16 non--zero eigenvalues) have 4 non--zero eigenvalues in common since they commute.}
 Since the pp--wave backgrounds obtained in this way always
 preserve at least 16 supersymmetries \cite{Blau:2002mw}
 we conclude that there is an additional  enhancement of supersymmetry
 by an extra 4 generators.\footnote{Note that
  this is  different  compared to
 the $AdS_5\times S^5$ or $AdS_4\times CP^3$ cases where  the number of supersymmetries
 remained the same.}

\subsubsection{BMN limit of the superstring action}

The quadratic part of the superstring action expanded  near
the classical   solution representing the relevant geodesic is
equivalent to the   superstring action for the  pp-wave background
(evaluated in the light-cone gauge).
In the pp--wave background discussed above the bosonic part of the superstring action
\eqref{action} becomes
\begin{equation}
S_B=-\frac{1}{2}T\int\Big[-4*dx^+dx^--{\mu^2}(r^2+z^2)\,*dx^+dx^++*dr dr+*dz dz+
*dy^{a'} dy_{a'}
\Big]\,.
\end{equation}
The fermionic part of the action is
\begin{equation}
S_F=-T\int\left(
i*dX^M\,\Theta\Gamma_M\mathcal D\Theta
-idX^M\,\Theta\Gamma_M\hat\Gamma \mathcal D\Theta
\right)\,.
\end{equation}
Fixing the light-cone kappa--symmetry  gauge
 $\Gamma^+\Theta=0$ simplifies the action
 considerably. For the type IIA case we find using (\ref{eq:IIAFslash})\footnote{The
  only non-zero components of the spin--connection are $\omega^{r-}\Gamma^r\Gamma^+$
  and $\omega^{z-}\Gamma^r\Gamma^+$ which do not contribute to the action in the
  gauge $\Gamma^+\Theta=0$.}
\begin{equation}
S_{F(IIA)}=-T\int\left(
-2i*dx^+\,\Theta\Gamma^-d\Theta
+2idx^+\,\Theta\Gamma^-\Gamma_{11} d\Theta
+2\mu*dx^+dx^+\,\vartheta\Gamma^-\gamma^7\Gamma^z\vartheta
\right)\,.
\end{equation}
For the type IIB case  (\ref{eq:IIBFslash}) leads to
\begin{equation}
S_{F(IIB)}=-T\int\left(
-2i*dx^+\,\Theta\Gamma^-d\Theta
+2idx^+\,\Theta\Gamma^-\sigma^3d\Theta
+2i\mu*dx^+dx^+\,\vartheta\Gamma^-\Gamma^r\Gamma_{(3)}\sigma^2\vartheta
\right)\,,
\end{equation}
where $\vartheta=\mathcal P_8\Theta$ and $\upsilon=(1-\mathcal P_8)\Theta$.
 Upon fixing the   light--cone gauge
\begin{equation}\label{lcg}
x^+=\frac{p^+\tau}{T}
\end{equation}
the action for the 8 transverse bosonic modes and the
fermionic fluctuations takes the form (upon a
 re--scaling of the fermionic fields)
 \begin{eqnarray}\label{lcga}
 S&=&-\frac{1}{2} T \int d\tau d\sigma \left[\partial_i r \partial^i r+\partial_i z \partial^i z+m^2(r^2+z^2)
 +\partial_i y^{a'} \partial^i y_{a'}
\right]\, \\
&&+iT\int d\tau d\sigma \left(
\vartheta^1\Gamma^-\partial_-\vartheta^1+
\vartheta^2\Gamma^-\partial_+\vartheta^2
+2m \,\vartheta^1\Gamma^-\Pi\vartheta^2+\upsilon^1\Gamma^-\partial_-\upsilon^1+
\upsilon^2\Gamma^-\partial_+\upsilon^2
\right)\, .\nonumber
 \end{eqnarray}
Here $\partial_{\pm}=\partial_\tau\pm \partial_\sigma$,\   $m\equiv \mu p^+/T$,  and
 $\vartheta^{1}$ and $\vartheta^{2}$ are
  two--component fermions  while  $\upsilon^1$ and $\upsilon^2$ are six--component fermions. The
  indices 1 and 2 indicate that these fermions originate from the two $D=10$ Majorana--Weyl
  spinors of the same (type IIB) or opposite (type IIA) chirality. The matrix $\Pi$ satisfies
   $\Pi^2=1$ and is equal to $-\Gamma^r$ and $i\Gamma^r\Gamma_{(3)}$
    in the type IIA and IIB cases respectively.

As was anticipated above,
the resulting physical   fluctuation spectrum  contains
 (i) the supercoset sector represented  by  two bosons $(r,z)$ and two
fermions $\vartheta^{1,2}$ of equal mass $m$,  and (ii)
the non--supercoset sector containing
$6$ massless bosons $y^{a'}$ and $6$ pairs of massless fermions $\upsilon^{1,2}$.
These two (2d supersymmetric)
sets of states are decoupled in the quadratic (pp-wave) approximation but
will start interacting  at higher orders of the near--BMN expansion
of the  superstring action.

\subsection{BMN limit along geodesic in $S^1 \times S^1 \subset S^2 \times T^6$ }\label{gpl}

Let us now   consider a more general  BMN   limit when the geodesic representing
the c.o.m. of the string runs along a ``diagonal'' direction in
the $S^1 \times S^1$  torus formed by
the equator $S^1 \subset S^2$ (with coordinate $\varphi$)
  and one of the
$S^1 \subset T^6$ directions, \emph{e.g.},  $y^4$. It
 can be  parameterized by the ``rotated'' coordinate
  $\varphi'$  as
\begin{eqnarray}\label{primebasis}
\varphi'&=&\cos\alpha\ \varphi+\sin\alpha\ \frac{y^4}{R}\,,\nonumber\\
y^{4'}&=&-\sin\alpha\ R\varphi+\cos\alpha\ y^4\,.
\end{eqnarray}
Here $\alpha$  is related to the ratio  of string angular momenta
along the two circles.\footnote{The corresponding classical string solution is:
$t= \kappa \tau, \ \varphi = p_\varphi \tau, \  y_4= p_{y} \tau $
with the Virasoro (massless geodesic) condition implying that \
$\kappa^2 = p^2_\varphi  +  R^{-2} p^2_{y}$ (see also Section 7).}
Setting  $\tilde x^{\pm}=\frac{t\pm\varphi'}{2}$ and zooming--in
 on the geometry seen by the particle
 by letting
\begin{equation}
x^+=\frac{\tilde x^+}{\mu}\,,\qquad x^-=\mu R^2\tilde x^-\,,\qquad \rho=\frac{r}{R}\,,\qquad\theta
=\frac{\pi}{2}+\frac{z}{R}\,,\qquad \mu R,\,\frac{R}{r},\,\frac{R}{z}\rightarrow\infty\,,
\end{equation}
we get from \rf{met}   the following pp-wave metric
\begin{equation}\label{gppw}
ds^2=-4dx^+dx^--\mu^2(r^2+\cos^2\alpha\,z^2)(dx^+)^2+dr^2+dz^2+dy_{a'}dy^{a'}\,,
\end{equation}
which generalizes our previous example  \rf{ppm}  to the case of $\alpha\not=0$.
Another special case is $\alpha= \frac{\pi}{2}$
when $\cos \alpha =0$ and the geodesic is running solely along $S^1 \subset T^6$.

The  fluxes of the type IIA solution \rf{RR}  become
\begin{eqnarray}
F_2&=&e^{-\phi}\mu\,dr\,dx^+\\
F_4&=&-\mu e^{-\phi}\cos\alpha\ dx^+\,dz\,J_2
-\mu e^{-\phi}(1-\cos\alpha)\,dx^+\,dz\,dy^5\,dy^{4'}\,,
\end{eqnarray}
where $J_2=dy^5dy^{4'}+dy^7dy^6+dy^9dy^8$.
This gives
\begin{equation}\label{slashF'}
\slashed F=-4i\mu\cos\alpha\ \gamma^7\Gamma^z\mathcal P'_8\Gamma^{+'}
-2i\mu(1-\cos\alpha)\gamma^7\Gamma^z\mathcal P_{16}\Gamma^{+'}
\end{equation}
where
\begin{equation}\label{p16}
\mathcal P_{16}=\frac{1}{2}(1-\Gamma^{6789}).
\end{equation}
 Note that in \eqref{slashF'} the projector $\mathcal P'_8$ is constructed with the
  gamma--matrices in the new basis corresponding to the change of variables performed
   in eq. \eqref{primebasis}, namely
 \begin{equation}\label{P8'}
 \mathcal P'_8=\frac{1}{4}\left(1+(\Gamma^{4'5}+\Gamma^{67}+\Gamma^{89})\Gamma^{4'56789}\right),
 \quad {\rm where} \quad \Gamma^{4'}=-\sin\alpha\Gamma^\varphi+\cos\alpha\Gamma^4
 \end{equation}
This  $\mathcal P'_8$   differs from the original
   $\mathcal P_8$ projector singling out the 8 supersymmetries of $AdS_2\times S^2\times T^6$:
    $\mathcal P'_8$ commutes with $\Gamma^{+'}=\frac{1}{2}(\Gamma^0+\Gamma^{\varphi'})$
   for all $\alpha\not=0$ while  $\mathcal P_8$ does not.
Using $\mathcal P'_8$ we may  again split  the  fermions $\Theta$
 into
 $\vartheta'$ and $\upsilon'$ which will now be linear combinations  of
  the original supercoset $\vartheta$
and non--supercoset $\upsilon$  fermions.

As usual in  the Penrose limit,
the 8 supersymmetries of $AdS_2\times S^2\times T^6$ get
 again  enhanced to 16 supersymmetries with
 the supersymmetry parameters satisfying $\Gamma^{+'}\epsilon=0$.
 However,  there is   no further  enhancement to 20 supersymmetries, unless $\alpha=0$.

 Now note that the projectors $\mathcal P_{16}$ and $\mathcal P'_8$ have the following property
$$
\mathcal P_{16} \mathcal P'_8=\mathcal P'_8.
$$
 Hence $\mathcal P_{16}$ acts as a unit matrix on
$\vartheta'=\mathcal P'_8\Theta$,  \emph{i.e. }
\begin{equation}\label{p16theta}
\mathcal P_{16}\vartheta'=\vartheta'\ ,
\end{equation}
and annihilates 16 of the 24 components of $\upsilon'=(1-\mathcal
 P'_8) \Theta $,  \emph{i.e. }
\begin{equation}\label{p16v}
\mathcal P_{16}\upsilon'=\tilde\upsilon'\ ,
\end{equation}
where $\tilde\upsilon'$ has only 8 non--zero components and therefore $\hat\upsilon'=(1-\mathcal P_{16})\upsilon'$ has 16 non--zero components.

We thus find that in the light-cone  kappa-symmetry gauge $\Gamma^{+'}\Theta=0$  the type IIA superstring action  takes  the form
\begin{eqnarray}\label{acgenp}
S_{IIA}&=&\hspace{-10pt}-T\int\Big[
-2*dx^+dx^--\frac{\mu^2}{2}(r^2+\cos^2\alpha\ z^2)\,*dx^+dx^++\frac{1}{2}(*dr dr+*dz dz+
*dy^{a'} dy_{a'}) \nonumber\\
&&
-2i*dx^+\,\Theta\Gamma^{-'}d\Theta
+2idx^+\,\Theta\Gamma^{-'}\Gamma_{11} d\Theta
\\
&&\hspace{-20pt}
+{\mu}(1+\cos\alpha)*dx^+\,dx^+\vartheta'\Gamma^{-'}\Gamma^z\gamma^7\vartheta'
+{\mu}(1-\cos\alpha)*dx^+\,dx^+\tilde\upsilon'\Gamma^{-'}\Gamma^z\gamma^7\mathcal{\tilde\upsilon}'
\Big]\ . \nonumber
\end{eqnarray}
In view of the properties \eqref{p16theta} and \eqref{p16v} of the  $\mathcal P_{16}$--projector,
it follows
 from  the action \eqref{acgenp}  that,  in the light--cone gauge $\Gamma'^{+}\Theta=0$, $\vartheta'$ and $\tilde\upsilon'$
  each describe  2  massive physical fermionic degrees of
   freedom while the remaining fermions
   $\hat \upsilon'$, satisfying
   $\mathcal P_{16}\hat\upsilon'=0$, represent  4 massless degrees
    of freedom.

To compute the masses
we again impose the
 light--cone gauge \eqref{lcg} and reduce the string action to the following  form
 describing only physical degrees of freedom
\begin{eqnarray}\label{lcgag}
 S&=&-\frac{1}{2} T \int d\tau d\sigma \Big(\partial_i r \partial^i r+\partial_i z \partial^i z+m^2\,r^2
 +m^2\cos^2\alpha\,z^2+\partial_i y^{a'} \partial^i y_{a'}
\Big)\, \nonumber\\
&&+iT\int d\tau d\sigma \Big[
\vartheta'^1\Gamma^{-'}\partial_-\vartheta'^1+
\vartheta'^2\Gamma^{-'}\partial_+\vartheta'^2
+m(1+\cos\alpha) \,\vartheta'^1\Gamma^{-'}\Pi\vartheta'^2\\
&&{}+\tilde\upsilon'^1\Gamma^{-'}\partial_-\tilde\upsilon'^1+
\tilde\upsilon'^2\Gamma^{-'}\partial_+\tilde\upsilon'^2+m(1-\cos\alpha) \,\tilde\upsilon'^1\Gamma^{-'}\Pi\tilde\upsilon'^2
+\hat\upsilon'^1\Gamma^{-'}\partial_-\hat\upsilon'^1
+\hat\upsilon'^2\Gamma^{-'}\partial_+\hat\upsilon'^2
\Big]\,,\nonumber
 \end{eqnarray}
where (as in \rf{lcga}) the fermions are appropriately re--scaled, $m=\mu p^+/T$ and the matrix $\Pi$ stands for $-\Gamma^r$
 or $i\Gamma^r\Gamma_{(3)}$, respectively, in the type IIA or IIB case.
This  action thus describes  two bosonic modes $r$ and $z$ with
 masses
 \begin{equation}\label{rz}
 m_r = m \ , \ \ \ \ \ \ \ \ \ \ m_z = m \cos\alpha \ ,
 \end{equation}\label{tv}
6 massless bosonic modes $y^{a'}$,
2 two--component fermions $\vartheta'^{1,2}$  with  mass $m_{\vartheta'}$ and
2 two--component fermions $\tilde\upsilon'^{1,2}$  with mass $m_{\tilde\upsilon'}$
\begin{equation}
m_{\vartheta'} = \frac{m}{2}(1 +  \cos\alpha)  \ , \ \ \ \ \ \ \ \ \ \
m_{\tilde\upsilon'} = \frac{m}{2}(1 - \cos\alpha)  \ ,
\end{equation}
as well as 4 massless fermionic  degrees of freedom
described  by  four--component  $\hat\upsilon'^{1,2}$.
As the  masses of bosonic  and fermionic degrees  of freedom  are different for
$\a\not=0$  there is no  effective 2d supersymmetry. The masses  still satisfy the 2d mass
sum rule
\begin{equation}
 \sum m_B^2-\sum m_F^2= m_r^2 + m^2_z - 2 m^2_{\vartheta'} - 2 m^2_{\tilde\upsilon'} =0        \ . \label{summm}
\end{equation}
This  is a manifestation of the UV finiteness of the corresponding GS action:
the bosonic masses originate from the curvature
 and the fermionic ones come from the RR background and the two are  related
for a  supergravity solution.

Let us note that  the  non--coset  fermions
$\upsilon'$  get non--trivial  masses due  to the non--decoupling of the broken--supersymmetry fermions from the
coset part of the theory (recall that $\upsilon'$ and $\vartheta'$ are a linear combination of the original coset fermions $\vartheta$
and non--coset $\upsilon$). This raises an issue of
 possible equivalence of the present GS  superstring theory and the
hybrid $AdS_2\times S^2\times T^6$ model of \cite{Berkovits:1999zq}
in which the $T^6$  part is  decoupled  from the supercoset part.\footnote{The hybrid model constructed
 in \cite{Berkovits:1999zq} consists of two completely
decoupled sectors, a free worldsheet $N=2$ superconformal theory on $T^6$ with
6 worldsheet spinor fermions  and a $\frac{PSU(1,
1|2)}{SO(1,1)\times U(1)}$ supercoset sigma--model (in addition to the  ghosts fields).
 In the hybrid model the
action for the supercoset sector differs from the GS supercoset action  \eqref{cosetL}
by an additional $J_1 J_3$  term  leading to a second--order kinetic term for the coset fermions $\vartheta'$. This additional term breaks
 kappa--symmetry but makes the worldsheet theory $N=2$ superconformal invariant.
 The definition of the  BMN limit involving a $T^6$ direction is not obvious in this case.
  It is not clear which condition on the coset fermions of the hybrid model may
  play the role of the light--cone kappa--symmetry gauge which reduced the GS
  superstring in the BMN limit to  a free theory. In addition, in the BMN
   limit the 6 non--supercoset worldsheet fermions of the hybrid model always remain
    massless, while two of their GS counterparts acquire a mass in the generic case
    (see eq.\eqref{lcgag}). Thus  the relation between the hybrid model of  \cite{Berkovits:1999zq}
      and the GS superstring
    on $AdS_2\times S^2\times T^6$, in which the $AdS_2\times S^2$ and $T^6$ sectors do not
    in general decouple, remains an  open issue even in the BMN limit.}


\section{Towards  exact solution of the quantum superstring  sigma-model}
\setcounter{equation}0

The string spectrum in the integrable AdS backgrounds, such as $AdS_5\times S^5$ or $AdS_4\times CP^3$
 can be found using Bethe--ansatz techniques. Above we provided evidence that the superstring sigma--model on $AdS_2\times
  S^2\times T^6$ is integrable, and we may thus expect that its spectrum is also
  described by a set of Bethe equations. Here we propose such a system for part of the
  spectrum that roughly speaking corresponds to the massive modes in the near--BMN expansion of Section~\ref{5.1},
   and is associated with the coset part of the supergeometry. As in the case of $AdS_3\times S^3\times
   T^4$  and similar
   backgrounds studied in \cite{Babichenko:2009dk}, the presence of the flat directions along $T^6$
   and their massless superpartners constitutes an obstacle for direct application of
    integrability methods, which work best for massive theories where one can define
    scattering states and the S--matrix. Nevertheless, we will propose a set of
    quantum Bethe equations, from which
    we will easily reconstruct
    the massive part of the BMN spectrum in Section~\ref{5.1}. The same equations cannot describe the BMN spectrum in Section~\ref{gpl} because of the mixing between coset and non--coset degrees of freedom in this latter case.

  Our starting point here is a restricted set of the equations
     of motion, with the $T^6$ directions and broken-supersymmetry fermions set to zero.
     As shown  in Section 4,
      this is a consistent truncation of the full theory.
      We can apply integrability methods to this restricted set of string configurations,
       and derive the classical version of the Bethe equations, which describe finite--gap
        solutions of the sigma-model
        \cite{Kazakov:2004qf} (see \cite{SchaferNameki:2010jy} for a review).
        A quantum counterpart of the finite-gap equations
	 can then be reconstructed using a structural analogy with the higher--dimensional AdS models
 \cite{Arutyunov:2004vx,Beisert:2005fw,Babichenko:2009dk,Zarembo:2010yz}.
 By construction, the resulting  quantum Bethe
  equations do not capture the massless $T^6$ modes of the string, but
   we conjecture that they correctly describe the massive sector of
    the spectrum, which may still form a closed subset of states due to integrability.\footnote{Let us
   draw  an analogy
   with    the   $AdS_5/CFT_4$ case: One can reconstruct the quantum
     Bethe ansatz for  the $su(2)$ sector  by studying classical strings restricted to
      the $\mathbbm{R}^1_{\rm time}\times S^3$ subspace of $AdS_5\times S^5$ \cite{Kazakov:2004qf}
       and then discretizing the resulting finite-gap equations \cite{Arutyunov:2004vx}.
       A restriction to a particular set of classical
       configurations does not make sense in quantum theory, but at the level of Bethe equations
       such a restriction is possible due to the
	underlying separation of variables.}

 The classical Bethe equations for the $PSU(1,1|2)/SO(1,1)\times U(1)$
 coset were derived in \cite{Zarembo:2010yz}. We shall first
 review this construction, and then propose a
  set of quantum Bethe equations that have the right classical limit and are
  structurally similar to the Bethe equations  for
  $AdS_5/CFT_4$ \cite{Beisert:2005fw} and  $AdS_4/CFT_3$ \cite{Gromov:2008qe}.

\subsection{Classical finite gap equations for the $PSU(1,1|2)/SO(1,1)\times U(1)$\\ supercoset model}

The starting point of the finite-gap method is the Lax--pair representation
 for the equations of motion. We do not understand at the moment how to include the Abelian
  directions along $T^6$ and the non--supercoset fermions (entering the Lax connection \eqref{Lax}) in the finite--gap integration scheme,
  and will thus concentrate on the
   $AdS_2\times S^2$ part of the supergeometry.\footnote{
  The Virasoro conditions  for the bosonic string  in $AdS_n \times S^n \times  T^{10-2n}$
  can be written as the vanishing of the sum of the stress tensor components
    $T_{\pm \pm} (AdS_n) + T_{\pm \pm} (S^n) + T_{\pm \pm} (T^{10-2n})=0$.
    Each of the three classical  stress tensors  is separately
    traceless and conserved and  thus $T_{\pm \pm}$ are functions of $\sigma \pm \tau$.
    In the case of $n=5$  when there is no toroidal part  we can make, say,
    $T_{\pm \pm} (AdS_5) = \mu^2$=const  by a  conformal transformation ($ \sigma \pm \tau \to f_\pm
    (\sigma \pm \tau)$),
     and then we will have
     $T_{\pm \pm} (S^5) = -\mu^2$  as a consequence of the Virasoro condition.
     The classical solutions  of the $AdS_5$ and $S^5$  sigma models satisfying  $T_{\pm \pm } =$const
       can then be described using the finite gap  construction  \cite{Kazakov:2004qf,Beisert:2005bm}.
      For $ n < 5$  for generic string motions  the stress tensor of the
       toroidal part is  non-zero. The remaining freedom of conformal transformations
       then  no longer allows us to make both $T_{\pm \pm} (AdS_n)$ and
       $T_{\pm \pm} (S^n)$   constant. We may make one of them constant or make
       their sum constant (by setting $T_{\pm \pm} (T^{10-2n})=\mu^2$)
   but then the standard finite  gap construction for the
   $ AdS_n$ and $S^n$   sigma models   will
   not directly apply, as the finite--gap solutions require $T_{\pm \pm } =$const
   for each of the two factors separately.
   The  classical  integrability  of strings in $AdS_n \times S^n \times  T^{10-2n}$
   should  allow one to construct    solutions  that  are more general than
   the finite  gap ones and  are  parameterized  by additional  holomorphic functions
   representing  generic  initial data (the values  of $T_{\pm \pm }$
   may be interpreted as specifying part of the initial data).
   This should be closely  related to an integrability-based  solution of the Cauchy problem
   for classical coset-space sigma models, which is an interesting open problem.}
    The string action then reduces to the
   $PSU(1,1|2)/SO(1,1)\times U(1)$ coset sigma--model \rf{cosetL}, to which we can apply the
   general procedure outlined in \cite{Babichenko:2009dk,Zarembo:2010yz}.

The equations of motion of the $PSU(1,1|2)/SO(1,1)\times U(1)$ sigma--model
follow from the flatness condition  for the Lax connection
\begin{equation}\label{lax}
 L_j=J_{j\,0}+\frac{{\tt x}^2+1}{{\tt x}^2-1}\,J_{j\,2}
 -\frac{2{\tt x}}{{\tt x}^2-1}\,\,\frac{1}{\sqrt{-h}}\,h_{jk}\varepsilon
 ^{kl}J_{l\,2}+\sqrt{\frac{{\tt x}+1}{{\tt x}-1}}\,J_{j\,1}
 +\sqrt{\frac{{\tt x}-1}{{\tt x}+1}}
 J_{j\,3}.
\end{equation}
The monodromy of the Lax connection defines the generating function for an infinite set of
conserved charges:
\begin{equation}
 \mathcal{M}({\tt x})={\rm P}\exp\left(
 \oint d\xi^jL_j(\xi ;{\tt x})
 \right).
\end{equation}
More precisely, the generating functions of the conserved charges are quasimomenta, the
eigenvalues of the monodromy matrix:
\begin{equation}\label{M(x)}
 \mathcal{M}({\tt x})=U^{-1} \exp\Big(p_l({\tt x})H_l\Big)\ U\ ,
\end{equation}
where $H_l$ are the Cartan elements of $\mathfrak{psu}(1,1|2)$.

Although the monodromy matrix
itself is an analytic function of the spectral parameter (except for an essential singularity
at ${\tt x}=0$), its eigenvalues are
not. In the defining supermatrix representation, $\mathcal{M}({\tt x})$ is a $4\times 4$ supermatrix. The Cartan generators are diagonal matrices, and thus to find quasimomenta we need to diagonalize $\mathcal{M}({\tt x})$. The eigenvalues of a $4\times 4$ matrix are solutions of an algebraic equation of degree four and are thus defined on an algebraic curve, a four--fold cover of the complex ${\tt x}$ plane. Indeed, if two eigenvalues coincide at some point ${\tt x}_*$, encircling this point in the complex plane will result in their permutation. Therefore, the eigenvalues of $\mathcal{M}({\tt x})$ may have branch points with the monodromy in the permutation group. It is convenient to give a more abstract, group-theoretic characterization of possible monodromies, in terms of quasimomenta.  The quasimomenta, by definition, parameterize the conjugacy class of the monodromy matrix viewed as a group element of  $PSU(1,1|2)$. The set of conjugacy classes of a (super)Lie group is its maximal torus divided by the Weyl group. However, the quasimomenta defined in (\ref{M(x)}) map the conjugacy class of $\mathcal{M}({\tt x})$ to the Cartan subalgebra of $\frak{psu}(1,1|2)$. The map from the maximal torus / Weyl group to the Cartan algebra is a multivalued map, and consequently $p_l({\tt x})$ are multivalued functions of ${\tt x}$. We may make $p_l({\tt x})$ single-valued on the complex plane with cuts $C_{l,i}$. As soon as we cross the cut $C_{l,i}$ (or, better, once we encircle one of its endpoints), the quasimomentum $p_l({\tt x})$ undergoes a shift by $2\pi n_{l,i}$ (because the set of conjugacy classes is a torus) and a transformation from the Weyl group $p_l({\tt x})\rightarrow p_l({\tt x})-A_{lm}p_m({\tt x})$ (because we need to mod out by Weyl transformations). Here $A_{lm}$ is the Cartan matrix of $\frak{psu}(1,1|2)$.

Additional constraints on the quasimomenta comes from the $\mathbbm{Z}_4$--symmetry of the coset, which acts on the Cartan generators of $\frak{psu}(1,1|2)$ as
\begin{equation}\label{defS}
 \Omega (H_l)=H_mS_{ml}.
\end{equation}
The constraints of analyticity and the  $\mathbbm{Z}_4$--symmetry result in the following integral representation for the quasimomenta \cite{Babichenko:2009dk}:
\begin{equation}\label{}
 p_l({\tt x})=-\frac{\kappa _l{\tt x}}{{\tt x}^2-1}
 +\int_{}^{}d{\tt y}\,
 \frac{\rho _l({\tt y})}{{\tt x}-{\tt y}}
 -S_{lm}\int_{}^{}\frac{d{\tt y}}{{\tt y}^2}\,\,\frac{\rho_m({\tt y}) }{{\tt x}-\frac{1}{\tt
 y}}\,,
\end{equation}
where the densities $\rho _{l,i}({\tt x})$ are defined on the cuts $C_{l,i}$ and are
determined by a set of singular integral equations:
\begin{equation}\label{classbethe}
 A_{lm}-\!\!\!\!\!\!\!\!\;\int_{}^{}d{\tt y}\,\frac{\rho _m({\tt y})}{{\tt x}-{\tt y}}
 -A_{lk}S_{km}\int_{}^{}\frac{d{\tt y}}{{\tt y}^2}\,\,
 \frac{\rho_m({\tt y}) }{{\tt x}-\frac{1}{\tt
 y}}=A_{lk}\,\frac{\kappa _k{\tt x}}{{\tt
 x}^2-1}+2\pi n_{l,i},\qquad {\tt x}\in C_{l,i}.
\end{equation}
The constants $\kappa _l$ could in principle be extracted from the semiclassical analysis of
the auxiliary linear problem for the Lax operator, but they are also constrained by group theory,
 which will be sufficient to determine them uniquely in our case. The constraints are:
\begin{equation}\label{eigenkappa}
 S_{lk}\kappa _k=-\kappa _l,
 \qquad
 \kappa _lA_{lk}\kappa _k=0.
\end{equation}

For the $PSU(1,1|2)$ coset \cite{Zarembo:2010yz},
\begin{equation}\label{}
 A=
\begin{pmatrix}
 0 & -1 & 0 \\
 -1 & \hphantom{-}2 & -1 \\
 0 & -1 & 0
\end{pmatrix},
\qquad
 S=
\begin{pmatrix}
 0& 0& -1 \\
 0&-1 &0 \\
 -1 &0 &0
\end{pmatrix},
\qquad
\kappa =\,{\rm const}\,\cdot \begin{pmatrix}
 1   \\
  0 \\
  1 \\
 \end{pmatrix},
\end{equation}
and the finite-gap integral equations take the form:
\begin{eqnarray}\label{fingap}
 2\pi n_{1,i}&=&\int_{}^{}d\y\,\rho_2 (\y)K(\x,\y)\ , \nonumber \\
\frac{2\kappa \x}{\x^2-1}+2\pi n_{2,i}&=&
2\pint_{}^{}d\y\,\rho_2 (\y)K(\x,\y)
-\int_{}^{}d\y\,\rho_1 (\y)K(\x,\y)-\int_{}^{}d\y\,\rho_{3} (\y)K(\x,\y)\ , \nonumber \\
2\pi n_{3,i}&=&\int_{}^{}d\y\,\rho_2
(\y)K(\x,\y)\ .
\end{eqnarray}
with
\begin{equation}
 K(\x,\y)=\frac{1}{\x-\y}+\frac{1}{\y^2}\,\,\frac{1}{\x-\frac{1}{\y}}\,.
\end{equation}
These  equations are schematically depicted in fig.~\ref{dynads2}.
\begin{figure}[t]
\centerline{\includegraphics[width=14cm]{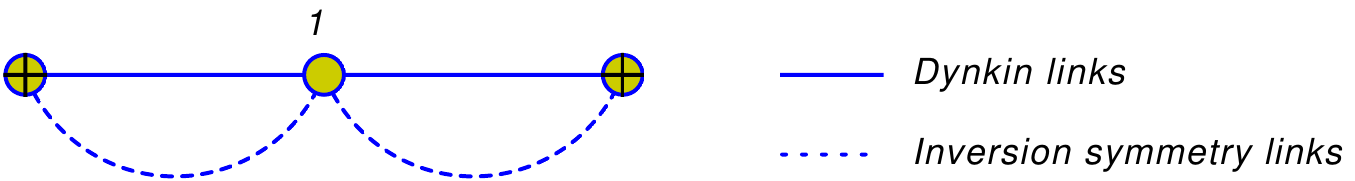}}
\caption{\label{dynads2}\small The Dynkin diagram for the Bethe equations.
The Dynkin links correspond to the first term in (\ref{classbethe}). The
inversion-symmetry links originate from the second term ($A\cdot S$), and
in our case are proportional to the Dynkin links.}
\end{figure}
The equation for the bosonic $\rho_2({\tt x})$ density alone (with $\rho_1$, $\rho_3$ set to zero) describes classical strings on $S^2\times \mathbbm{R}_{\rm time}$ \cite{Beisert:2004ag}, as it should.

The finite--gap equations characterize classical solutions of the sigma-model in terms of the spectral data for the Lax operator. The inverse-scattering transformation allows, in principle, to reconstruct any given solution from its spectral data, but for the most part we are interested in the conserved charges that the solution carries, and those can be computed from the quasimomenta.

\subsection{Proposal for asymptotic Bethe ansatz
 equations \\ for the supercoset sector of states}

We can now quantize the above classical Bethe equations
 using the rules of \cite{Arutyunov:2004vx,Beisert:2005fw}, formulated for general
  $\mathbbm{Z}_4$--cosets in  \cite{Zarembo:2010yz}. In quantum theory,
  the cuts  are composed of discrete Bethe roots, that satisfy a set of functional equations, the Bethe equations. The finite-gap equations constitute their classical limit, in the situation when the number of roots
  is large, and the distance between nearby roots is small such that
  their distribution can be characterized by a continuous density.

  For strings in $AdS_5\times S^5$, it was possible to reconstruct the  quantum
  Bethe equations from their classical counterpart with the help of extra
  information from the underlying spin chain  \cite{Beisert:2005fw}. The same
  approach worked also for the $AdS_4\times CP^3$ background \cite{Gromov:2008qe}.
  In the present case the spin chain description of the dual CFT is not known, but
   one may observe that the set of rules to convert classical finite--gap equations
   to quantum Bethe equations follows a uniform model-independent pattern. The
    same pattern arises in the case at hand, since the structural elements that appear in
     the finite-gap equations are similar. We may thus apply the same set of rules to convert the classical finite-gap equations into the quantum Bethe equations.

The basic variable of quantum Bethe ansatz is the rapidity $u$, which is related
to the spectral parameter by the Zhukowsky transformation
\begin{equation}
 \x+\frac{1}{\x}=u.
\end{equation}
The dependence on the sigma-model coupling enters through the shifted variables
\begin{equation}\label{Zhuk}
 \x^\pm+\frac{1}{\x^{\pm}}=u\pm i\hbar,
\end{equation}
where ``the Planck constant" $\hbar$ is a function of the sigma-model coupling,
$ (T R^2)^{-1} = \frac{2\pi \alpha'}{R^2}$
(with $R$ being the AdS radius)  which
is  not determined by the integrability of the model;
we can only say that  $\hbar$ coincides
 with the coupling up to possible higher--order $\alpha'$
 corrections, i.e.  $\hbar={2\pi \alpha '\over R^2}  +O(\alpha'^2)$.

The Zhukowsky variables determine the momentum and energy of a single worldsheet excitation:
\begin{equation}
 \,{\rm e}\,^{i\hbar p}=\frac{\x^+}{\x^-}\,,\qquad
 \hbar\varepsilon =\frac{i}{\x^+}-\frac{i}{\x^-}+1.
\end{equation}
The rapidities of individual excitations in a multiparticle state satisfy a set of Bethe equations:
\begin{eqnarray}\label{Bethe}
 1&=&\prod_{k}^{}\frac{\x_{1,j}-\x_{2,k}^+}{\x_{1,j}-\x_{2,k}^-}\,\,
 \frac{1-\frac{1}{\x_{1,j}\x_{2,k}^-}}{1-\frac{1}{\x_{1,j}\x_{2,k}^+}}\ ,
\nonumber \\
\left(\frac{\x_{j,2}^+}{\x_{j,2}^-}\right)^L
&=&
\prod_{k\neq j}^{}
\frac{\x^+_{j,2}-\x^-_{k,2}}{\x^-_{j,2}-\x^+_{k,2}}\,\,
\frac{1-\frac{1}{\x^+_{j,2}\x^-_{k,2}}}{1-\frac{1}{\x^-_{j,2}\x^+_{k,2}}}\,
\sigma ^4_{\rm BES}(u_{j,2},u_{k,2})
\nonumber \\
&&\times
\prod_{k}^{}\frac{\x^-_{2,j}-\x_{1,k}}{\x^+_{2,j}-\x_{1,k}}\,\,
\frac{1-\frac{1}{\x^+_{2,j}\x_{1,k}}}{1-\frac{1}{\x^-_{2,j}\x_{1,k}}}\,
\prod_{k}^{}\frac{\x^-_{2,j}-\x_{3,k}}{\x^+_{2,j}-\x_{3,k}}\,\,
\frac{1-\frac{1}{\x^+_{2,j}\x_{3,k}}}{1-\frac{1}{\x^-_{2,j}\x_{3,k}}}\ ,
\nonumber \\
1&=&\prod_{k}^{}\frac{\x_{3,j}-\x_{2,k}^+}{\x_{3,j}-\x_{2,k}^-}\,\,
 \frac{1-\frac{1}{\x_{3,j}\x_{2,k}^-}}{1-\frac{1}{\x_{3,j}\x_{2,k}^+}}\ .
\end{eqnarray}
The BES/BHL scattering phase $\sigma _{\rm BES}(u,v)$
\cite{Beisert:2006ez,Beisert:2006ib} (see \cite{Vieira:2010kb} for a review) is a
fairly complicated function of the rapidities. It admits the following integral representation
\cite{Dorey:2007xn}:
\begin{eqnarray}
 \sigma _{\rm BES}(u,v)&=&\exp\left(-i\sum_{r,s=\pm}^{}\chi (\x^r,\y^s)\right),
\nonumber \\
 \chi (\x,\y)&=&\frac{i}{4\pi ^2}\oint_{|z|=1,|w|=1}
 \frac{dz\,dw}{(\x-z)(\y-w)}\,\,\ln
 \frac{\Gamma\left(1+\frac{i}{2\hbar}\left(z+\frac{1}{z}-w-\frac{1}{w}\right)\right)}
 {\Gamma\left(1-\frac{i}{2\hbar}\left(z+\frac{1}{z}-w-\frac{1}{w}\right)\right)}\,.\label{pha}
\end{eqnarray}

The above Bethe  equations are supposed to describe the spectrum of the string in the light-cone
 gauge defined by the geodesic of a massless particle spinning around $S^2$.
  The parameter $L$ that enters the left-hand-side of the equations is the
   angular momentum of the ground state ($L=J$). Once  $\x^\pm _{2,k}$ are determined,
   the light-cone energy of the string
   states for a given collection of Bethe roots is
   computed by\footnote{The full AdS energy
   contains also the ``vacuum''  $L=J$ term.}
\begin{equation}\label{Benegry}
 E= \frac{1}{\hbar}\sum_{k}^{}\left(\frac{i}{\x_{2,k}^+}-\frac{i}{\x_{2,k}^-}+1\right).
\end{equation}
The physical states should satisfy the level matching (zero-momentum) condition:
\begin{equation}\label{Bmomentum}
 \prod_{k}^{}\frac{\x_{2,k}^+}{\x_{2,k}^-}=\,{\rm e}\,^{i\hbar P_{\rm tot}}\equiv 1.
\end{equation}
Only the roots on the middle node of the Dynkin diagram ($u_2$ roots) carry energy and momentum.
 The other two types of roots are auxiliary. They just change the flavor composition of the state.
 These roots are fermionic in the sense that adding an odd number of $u_1$ and $u_3$ roots produces a fermion.

We should stress that the above Bethe ansatz equations have not been derived from first principles but rather conjectured, building upon structural analogy with other AdS backgrounds.
It is a straightforward, albeit lengthy,  exercise to show that
in the classical limit $\hbar\rightarrow 0$,
  the quantum Bethe equations (\ref{Bethe}) reduce to the
  classical finite-gap equations (\ref{fingap}),
  with the densities  defined by
\begin{equation}
 \rho _{l}(\x)=2\hbar\sum_{k}^{}\frac{\x_{l,k}^2}{\x_{l,k}^2-1}\,\delta \left(\x-\x_{l,k}\right).
\end{equation}
Hence the conjectured equations correctly capture the classical spectrum of the string (this however by construction). It would be important to test them beyond the classical approximation.

Let us check that the Bethe equations (\ref{Bethe}) reproduce the  BMN spectrum from sec.~\ref{5.1}. Their simplest
solution is a solitary $u_2$ root. According to (\ref{Benegry}), (\ref{Bmomentum}) and (\ref{Zhuk}),
at $\hbar\rightarrow 0$ it describes a worldsheet excitation with
the momentum and energy
\begin{equation}
 p=\frac{2\x_2}{\x^2_2-1}\,,\qquad
 \varepsilon =\frac{\x^2_2+1}{\x^2_2-1}\,,
\end{equation}
which in fact parameterize  the dispersion relation of a relativistic particle with mass $m^2=1$. The Bethe
equations reduce to the quantization condition for the momentum: $p=2\pi n/\hbar L$. This is consistent
with the light-cone quantization, in which the length of the string is given by its centre-of-mass
 momentum in the target space:
${\rm Length}=2\pi \alpha 'L/R^2=L\hbar$.

The Bethe equations admit solutions with the $u_1$ and/or $u_3$ roots added to the solitary $u_2$ root.
The Bethe equations determine the positions of the auxiliary roots:
\begin{equation}
 \x_{1,3}=\frac{\x_{2}^++\x_2^-}{\x_2^+\x_2^-+1}\approx \frac{2\x_2}{\x^2_2+1}\,.
\end{equation}
These roots change the quantum numbers of the string state without changing its energy. The $1-2$
 and $2-3$ complexes are fermions and the $1-2-3$ complex is a boson. The single type--2 root
 describes the transverse mode of the string on $S^2$, the $1-2-3$ stack describes the $AdS_2$ mode,
 and the two fermionic solutions correspond to the two coset fermions that survive the kappa-symmetry
 gauge fixing. We thus reconstruct the $2_b+2_f$ massive BMN modes but are obviously
 missing the $6_b+6_f$ massless modes, which correspond to
 the string
 fluctuations in the $T^6$ directions and their superpartners. At the moment we do not understand
 how to include these modes in the Bethe ansatz framework.

\section{Remarks on semiclassical
 strings
 in $AdS_2 \times S^2 \times T^6$}\label{ABC}
\setcounter{equation}0

As in the $AdS_5 \times S^5$ case,  one may try to check the above
Bethe--ansatz  based solution for the
string  spectrum against direct superstring predictions in the limit of large
(semiclassical) values of charges.

Already at   the string tree level  one may compare the  corresponding
(Landau--Lifshitz--type) effective
action  following, for coherent states, from the  ``one-loop''
spin chain Hamiltonian formally associated to the Bethe ansatz  in \rf{Bethe}
to the corresponding ``fast-string'' ($J \gg 1$)
 limit of the string  action.
The two are known to match \cite{Kruczenski:2003gt} in the $AdS_5 \times S^5$ case due to
a special non-renormalization property of the leading order  correction in the large  $J$ limit.

For example, let us  consider  classical strings  moving in the $S^2$ subspace.
  Following the discussion in Section 5.2 of \cite{Kruczenski:2004cn}  and eliminating ``fast''
 angular coordinates from the phase-space string action on
$R_t \times S^2$ one finds that it reduces to the same   Landau--Lifshitz model
that is found, \cite{Kruczenski:2003gt}, for ``slow'' coordinates of
a fast-moving string  in $R \times S^3$.
In the  $R_t \times S^2$  case, however, the two degrees of freedom are the
 coordinate and momentum of the
``slow'' degree of freedom  of $S^2$. The same action is known also to emerge  from the
ferromagnetic XXX$_{1/2}$    spin  chain Hamiltonian (the 1--loop  Hamiltonian in the
 $su(2)$ sector of $\cal N$=4 SYM theory).

 This coincidence  is, in fact, in agreement with the structure   of  the Bethe ansatz
 proposed in the previous section:
 In the formal  weak--coupling limit ($\hbar= { 2 \pi\alpha' \over R^2 }+ ... \to \infty)$
  the central--node  part of the
  Bethe equations \rf{Bethe} takes the same form as for
   the  $su(2)$ Heisenberg spin chain (or rather $SO(3)$   subsector of the one-loop
   $SO(6)$ spin   chain):\footnote{Here the  role of the second spin component of the $su(2)$ sector
   or number of impurities
    is played by the string oscillation number.}
      $u_j$ and ${\rm x}_j$ are then of order $\hbar$,\ ${\rm x}
   ^{\pm}_j = u_j \pm   i\hbar$, the BES phase disappears, and
only the first term on the r.h.s. of the second  equation in \rf{Bethe} contributes.
   This implies that  the same Landau-Lifshitz
   model should indeed  emerge   as a description of the coherent  long-wave-length  spin-wave
   states.\footnote{Here we  compare (i)  Landau-Lifshitz model that comes out of the
string  phase-space classical action after isolating a fast degree of freedom
and  (ii) Landau-Lifshitz model that comes
out of a spin chain  Hamiltonian that is suggested to exist  by
the form of the postulated BA, assuming we formally
take
the weak coupling  limit there. The matching is then exactly as in  the $AdS_5 \times S^5$ case
 where  we  may  consider $O(3)$ part of the  1-loop  $O(6)$  subsector of
spin chain Hamiltonian, and the  matching is effectively
due  to supersymmetry  protection. It is unclear how such Hamiltonian
may come out of the dual CFT in the present $AdS_2\times S^2$ case.}

\

At the one-loop string level
one may  carry out  a  comparison between  the string theory predictions for the
1-loop corrections to  energies of semiclassical   strings moving in the $AdS_2 \times S^2$ part
 and the predictions
of the Bethe ansatz equations  \rf{Bethe}. One may attempt to do  this in general
 following the approach used in the $AdS_5 \times S^5$ and   $AdS_4 \times CP^3$ cases
 \cite{Gromov:2007cd,Gromov:2007ky,Gromov:2008ec}.
This should provide, in particular,  a check of the BES phase factor in \rf{Bethe}.

It  is useful also to look at specific examples
of  semiclassical quantization of the simplest    string solutions.
Let us  consider strings moving only in the  $AdS_2 \times S^2$ part with the metric
given in \rf{met}.
The corresponding classical solutions can be reconstructed, via Pohlmeyer reduction \cite{Grigoriev:2007bu,
Hoare:2009rq},
  from solutions of the $N=2$
supersymmetric  sine-Gordon  theory (whose bosonic part is the direct  sum of the sine-Gordon and
sinh-Gordon models).
The standard ``vacuum'' (BMN) solution, \emph{i.e.} the  massless  geodesic wrapping a
 big circle of $S^2$  is $t= \varphi = \nu \tau$, \
$\nu = { J \over T R^2 },\  E= J $. It leads (as discussed in Section 5.1)
to the small--fluctuation spectrum
containing 2  massive  ($m = \nu$) bosonic modes, 2 massive  fermionic  modes and
6+6 massless modes.

In $AdS_2 \times S^2$  there
  is no place for  rigid circular string solutions described by rational
functions  so the next class of solutions in terms of simplicity  are rigid spinning or pulsating  strings
described by  elliptic  functions.
One  example is the giant magnon \cite{Hofman:2006xt} (an open string  spinning in
 $S^2$ with its ends on a big circle)  which  of course  has   the same classical  dispersion relation
as in the $AdS_5 \times S^5$ case. The corresponding
  small fluctuation spectrum can be analysed
  as in \cite{Minahan:2007gf,Papathanasiou:2007gd}, now starting  with  the
  $AdS_2 \times S^2 \times T^6$ superstring action.
  Instead of 4  massive bosonic fluctuations  from $AdS_5$  one finds
  one  from $AdS_2$  with the  same stability angle $\nu_1$;
instead of 4 bosonic fluctuations  from $S^5$   one gets 1
from $S^2$  with  the same stability angle $\nu_2= \nu_1 + 2 {\rm arccot}\ k$
(where $k$ is the spatial 2d momentum number);
 instead of 8 massive fermionic  modes  one gets  2 massive fermions from the supercoset part of the  action
with $\nu_3= \nu_1 +  {\rm arccot}\ k$.
 In addition,  there are  6  decoupled massless $T^6$
modes and  6 massless fermions (after $\kappa$-gauge fixing).
The ``non-coset'' fermions  do not get mass from the RR coupling
term in the  quadratic fermionic action
(due to the structure of the projector in \rf{FIIA} or \rf{IIBB})
while the induced connection term in the covariant derivative \rf{E} can be
rotated away.
As a result, the 1-loop correction to the   string energy  (determined by
$\nu_1 + \nu_2 - 2 \nu_3$)  vanishes
  as in the $AdS_5 \times S^5$ case \cite{Papathanasiou:2007gd}.

Another simple  elliptic solution is a  folded  string rotating around its center of mass in
$S^2$ \cite{Gubser:2002tv}:
\begin{eqnarray}  && t= \kappa \tau,\ \  \ \theta= \theta(\s), \ \ \
\varphi= w \tau ,\ \ \ \ \    \theta'^2 = \kappa^2 - w^2  \sin^2 \theta \ ,\\ &&
\sin \theta = \sqrt q\  {\rm sn} ( w \s | q ) \ , \ \ \ \ q=\sin^2 \theta_0= {\kappa^2 \ov w^2}\ ,
\ \ \ \  w= { 2 \ov \pi} {\rm K}(q) \ ,  \label{kp}
\end{eqnarray}
where ${\rm K}$ is the elliptic integral.
In the case of $AdS_n \times S^n\times T^{10-2n} $ (with $n=5,3,2$)  the
corresponding quadratic  fluctuation spectrum  can be found by
imposing  the static gauge on the fluctuations and turns out
 to be a simple truncation of the $AdS_5 \times S^5$ spectrum found in \cite{Beccaria:2010zn}.
It is described by a combination of  massive  bosonic
 and fermionic 2d  modes  with the following degeneracy $\times$ (mass)$^2$
\begin{eqnarray}
&& {\rm Bosons}: \ \ \ \ AdS_n: \ \ \   (n-1) \times \kappa^2 ; \nonumber \\
&&\ \ \ \ \ \ \ \ \ \ \ \ \ \ \ \ \ \
S^n: \ \ \  1 \times \kappa^2 [ 1-  { 2(\kappa^2 -w^2) \ov \kappa^2 - w^2  \sin^2 \theta}
 ]; \ \  (n-2)\times  ( 2 w^2 \sin^2 \theta -
\kappa^2)  \nonumber \\  &&\ \ \ \ \ \ \ \ \ \ \ \ \ \ \ \ \ \  \    T^{10-2n}: \ \ \  (10-2n) \times 0
\nonumber \label{bo}\\
&& {\rm Fermions}: \ \ \ \      2 (n-1) \times  w^2 \sin^2 \theta; \ \  (10-2n) \times 0
\ . \label{fenccc}\end{eqnarray}
The resulting mass sum rule  that checks the  1-loop  UV finiteness of the  GS string in the
static gauge is then universal in $n$  (see \cite{Beccaria:2010zn})
\be
\sum ( m^2_B  - m^2_F) = 2 \Big( \kappa^2 - w^2  \sin^2 \theta - \kappa^2
{ \kappa^2 -w^2 \ov \kappa^2 - w^2  \sin^2 \theta}\Big) = \sqrt{ - g} R^{(2)} \ .
\ee
Here   the coset part   decouples from the torus part of the model.
This is due, in particular, to the possibility of rotating away the connection
in the covariant derivative in the fermionic part, i.e.   the masses   come solely
from the  RR flux term coupling  in the quadratic fermionic action.
The resulting  1-loop correction to  the folded
string energy in  $AdS_2 \times S^2\times T^6 $ can be computed  by
direct combination of the expressions given  in \cite{Beccaria:2010zn}.

A similar  analysis can be carried out  for the circular
 pulsating solution  on $S^2$ \cite{Minahan:2002rc}  which  is described in conformal gauge by
 (cf. \eqref{kp})
 \begin{eqnarray}
 && t= \kappa \tau,\ \  \ \theta= \theta(\tau), \ \ \
\varphi= m \sigma ,\ \ \ \ \    \dot \theta^2 = \kappa^2 - w^2  \sin^2 \theta \ ,\\ &&
\sin \theta =\sqrt q \  {\rm sn} ( m \tau | q ) \ , \ \ \ \ \ \ \
q=\sin^2 \theta_0= {\kappa^2 \ov m^2}\ , \ \  \ \ \ \  w= { 2 \ov \pi} {\rm K}({\kappa \ov
w}) \ .
\label{lp}
\end{eqnarray}
For a superstring  in  $AdS_n \times S^n\times T^{10-2n} $ ($n=5,3,2$)  the
corresponding  quadratic fluctuation spectrum  in the static gauge
is again a  truncation of the $AdS_5 \times S^5$ spectrum in \cite{Beccaria:2010zn}:
\begin{eqnarray}
&& {\rm Bosons}: \ \ \ \ AdS_n: \ \ \   (n-1) \times \kappa^2 ; \nonumber \\
&&\ \ \ \ \ \ \ \ \ \ \ \ \ \ \ \ \ \
S^n: \ \ \  1 \times \kappa^2 ( 1- {2  \ov  \sin^2 \theta}
 ); \ \ \  (n-2)\times  ( \kappa^2 - 2 m^2 \sin^2 \theta)  \nonumber \\  &&\ \ \ \ \
 \ \ \ \ \ \ \ \ \ \ \ \ \  \    T^{10-2n}: \ \ \  (10-2n) \times 0
\nonumber\label{bod}\\
&& {\rm Fermions}: \ \ \ \      2 (n-1) \times  (\kappa^2 - m^2 \sin^2 \theta); \ \ \  (10-2n) \times 0
\ . \label{fye}
\end{eqnarray}
Again, we get a  universal  mass sum rule that checks the UV finiteness  (see \cite{Beccaria:2010zn})
\begin{eqnarray}
\sum ( m^2_B  - m^2_F) = 2 \Big( m^2 - {\kappa^2 \ov  \sin^2 \theta }\Big) = \sqrt{ - g} R^{(2)}\,.
\end{eqnarray}
As in the previous examples,   the masses   come only from the RR coupling term  and the
toroidal part decouples\footnote{From the algebraic curve point of view the 1--loop fluctuation
   frequencies will come only   from the supercoset part.}  and
does not contribute to the 1--loop correction to  the string energy.

 Given that the  Bethe  ansatz  equations \rf{Bethe}
 were  constructed  on the  basis of the supercoset part of
 the  string model  (i.e. under the assumption of the decoupling of
 the non--supercoset parts)
   this  guarantees the agreement  between their predictions   and the
  direct superstring predictions for the 1--loop energies of
   the above solutions.
  This agreement  should  also extend  to the leading strong coupling
   finite size  (TBA)  generalization
  of the asymptotic Bethe ansatz  equations  as there should be a direct
   analog of the   analysis in \cite{Gromov:2009tq}.
  One may also hope
  that higher order (2-loop,  etc.) string corrections to energies of semiclassical string
  solutions  in  $AdS_2 \times S^2$
  will not be sensitive to contributions of  massless toroidal modes.
  A priori, this may not apply to the full quantum string spectrum
  which  may be sensitive to finite-size corrections due to massless modes.

\

Let us now comment on   the case of semiclassical  strings moving also  in the toroidal part.
In general, the radii $\rr_i$   of $T^6$ (or, more generally,
 its constant metric) are  free parameters of the model,
in addition to the radius $\rR$ of  $AdS_2$ and $ S^2$. Thus  by varying  $\rr_i/\rR  $
 we may suppress or enhance the contributions of configurations
 in which the  string is moving  in $T^6$.
 For example, we may generalize the above solutions to the case when the c.o.m.
   of the string moves in  $S^1$ of $T^6$, \
   $y \equiv \rr \psi, \  \psi = p  \tau, \ \
    P = { r^2 \ov  \a'} p.$
 That will  generalize  the energy relation $E=J$ we  had  for   the BMN  state  in
 $S^2$ as follows:
 \begin{equation} \kappa^2 = \nu^2 +  \gamma^2 p^2 , \ \ \ \ {\rm  i.e. } \ \ \ \
 E^2 = J^2  +   \gamma^{-2}  P^2 , \ \ \ \ \ \  \gamma \equiv  { {\rr } \ov \rR} \ .
 \end{equation}
 The corresponding   fluctuation spectrum was already discussed in Section 5.2.
 While the bosonic modes from $T^6$  remain  massless
 2 of the non--supercoset--like fermions $\upsilon'$  get non--zero mass,
 i.e. the coset and non--coset sectors no longer decouple.

 This non--decoupling will happen also for  extended string solutions, e.g.,
 a simple
  circular spinning string solution  constructed by allowing the string to
 wind around a big circle of $S^2$ as well as around a  circle in $T^6$
 ($n$ and $k$ are integer winding numbers)
\begin{eqnarray}
&&t= \kappa \tau\ , \ \ \theta= { \pi \ov 2} \ , \ \ \  \varphi= w \tau + n \sigma, \ \ \ \
\ \ \psi = p \tau -   k \s \ , \nonumber \\
&& \kappa^2 = w^2 + n^2  +  \gamma^2 ( p^2 + k^2)  \ , \ \ \ \ \    w m  =\gamma^2 p k \ .
\end{eqnarray}
Here again the non-coset fermions  will get non-trivial masses
and give a non-trivial contribution to the 1-loop correction to the energy.

Imposing the  conformal gauge,  the general $T^6$ solution
can be written as $y_{m'} = u_{m'}(\sigma_+) +\tilde  u_{m'}(\sigma_-), \ \
\sigma_\pm = \sigma \pm \tau$, so that
$T_{++}(T^6)  = u'^2(\sigma_+), \ T_{--}(T^6) =\tilde  u '^2(\sigma_-)$.
We may fix the residual conformal diffeomorphisms  by assuming that
$T_{++}(AdS_2)  =T_{--}(AdS_2) = -\mu^2=$const.
Then the Virasoro conditions $T^{(tot)}_{\pm \pm}=0$ imply that
\begin{eqnarray}
T_{++}(S^2) = (\mu^2 - u'^2) \equiv h^2(\sigma_+)\ , \ \ \ \ \ \ \
T_{--}(S^2) = (\mu^2 - \tilde u'^2) \equiv \tilde h^2(\sigma_-) \ .
\end{eqnarray}
Since the $T_{\pm \pm }(S^2)$  components  are now non-constant,
as was already mentioned  in Section 6,  in the case
when the string moves in $T^6$    one cannot describe the  corresponding
$AdS_2 \times S^2 \times T^6$ solutions  using  the finite gap  construction.

Let us mention that a related  problem appears also
  when one tries  to apply the Pohlmeyer reduction approach to
describe such string  solutions.
Instead of the  sinh-Gordon theory plus the standard  sine-Gordon  theory
as found in the $AdS_2 \times S^2$ case
one now  ends up with
\begin{eqnarray}
L=  \partial_+ \chi \partial_- \chi
   -  {\mu^2 \ov 2 } \cosh2 \chi  + \partial_+ \varphi \partial_- \varphi    +
    {1 \ov 2 } h(\sigma_+)\tilde h(\sigma_-) \cos 2 \varphi \ .
\end{eqnarray}
In the equation for $\varphi$  one can  formally  replace
 $h(\sigma_+)\tilde h(\sigma_-)$  by a constant  performing
  a conformal redefinition of
 $\sigma_\pm$. The result will be  the   sine--Gordon
 equation on a complicated 2d domain, determined by functions
 that parameterize  the $T^6$ solution.

\section{Conclusion}

In this paper we considered  GS superstrings   in  $AdS_2\times S^2
\times T^6$ backgrounds
that represent consistent embeddings of $AdS_2\times S^2$ space-time
into critical  string theory.
We have shown that the full 10d structure of the superstring theory
cannot be captured
just  by its   effective $AdS_2 \times S^2$  ``supercoset'' part
described  by  an
 integrable  sigma--model on $PSU(1,1|2)/SO(1,1)\times U(1)$,  though
 the latter is, indeed,  a  consistent classical  truncation of the full theory.
 Nevertheless, we have provided direct  evidence for the classical
integrability of
 the complete theory by constructing the Lax representation of its
equations of motion to the second order in fermions.

In general,  the $T^6$ sector  does not decouple from the  $AdS_2
\times S^2$  one
due to  non-zero  RR flux  components along the  $T^6$ directions
which  couple derivatives  of the $T^6$
 coordinates to the GS fermions.  We illustrated this in the example
of the second BMN limit in  Section 5.2.
Still, for strings moving only in  the  $AdS_2\times S^2$ subspace
the decoupling does  take place at
the one-loop  level (Section 7)  and that  may extend to all orders
in the semiclassical (large-charge) expansion.

For the part of the string spectrum which is associated with the
supercoset sector
 we have proposed Bethe equations that are of the asymptotic Bethe
ansatz type.
 They are supposed to describe quantum states with sufficiently large
quantum numbers up to exponential corrections. In other words, they
describe the string sigma--model on an infinite plane rather than on a
cylinder (the length of the string in the light-cone gauge is
proportional to its light--cone momentum, and if the light--cone
momentum is large, the internal length of the string goes to
infinity). Application of a more general TBA or Y--system framework
(see \cite{Gromov:2010kf} for a review) to this theory will likely
require understanding the massless $T^6$ modes and the non--coset
fermionic excitations of the string
 that we have ignored in our analysis of the Bethe equations.

The super $AdS_2\times S^2$ background is one of the few GS--type
cosets, which are simultaneously integrable and conformal. Other
possible candidates that contain an $AdS_2$ factor are $AdS_2\times
S^2\times S^2$ and $AdS_2\times S^3$  \cite{Zarembo:2010sg}. It would
be interesting to find critical string backgrounds which contain these
cosets as consistent truncations. It would be also of interest to verify if the fermionic T-duality of the supercoset model \cite{Adam:2009kt} persists in the presence of the non-supercoset fermions.

\section*{Acknowledgments }
The authors are grateful to R. Roiban for useful comments on the draft of this paper.
D.S. is thankful to N. Berkovits for the discussion of the hybrid model.
A.A.T. would like to thank G. Arutyunov,
N. Gromov,  B. Hoare, R. Roiban   and B. Vicedo  for useful
discussions. Work of D.S. was partially supported by the INFN
Special Initiative TV12, by an Excellence Grant of Fondazione Cariparo
(Padova) and the grant FIS2008-1980 of the Spanish Ministry of Science
and Innovation. The research of L.W. is supported in part by
NSF grants PHY-0555575 and PHY-0906222. The work of K.Z. was supported in part by the Swedish
Research Council
under contract 621-2007-4177, in part by the ANF-a grant 09-02-91005,
in part by the RFFI grant 10-02-01315, and in part
by the Ministry of Education and Science of the Russian Federation
under contract 14.740.11.0347.


\def\thesection{}
\def\theequation{A.\arabic{equation}}\label{A}
\section{Appendix A. Main notation and conventions}
\setcounter{equation}0

We assume the  metric to have  the `almost plus' signature
$(-,+,\cdots,+)$.
Generically, the tangent space vector indices are labeled by letters from the
beginning of the
Latin alphabet,  while  letters from the middle of the Latin alphabet stand
for curved (world)
indices. The spinor indices are labeled by Greek letters.
The direct product  $AdS_2\times S^2\times T^6$  is parametrized, respectively, by the coordinates $x^m$ $(m=0,1)$, $x^{\hat m}$ $(\hat m=2,3)$ and $y^{m'}$ $(m'=4,5,6,7,8,9)$. Its vielbeins are, respectively, $e^{ a}=dx^{m}\,e_{m}{}^{a}(x)$  (${a}=0,1$),
$e^{\hat a}=dx^{\hat
m}\,e_{\hat m}{}^{\hat a}(\hat x)$ (${\hat a}=2,3$) and $e^{a'}(y)=dy^{a'}$.
The $AdS_2$ curvature is
\be\label{ads2c}
R_{  a  b  c}{}^{  d}=\frac{2}{R^2}\,\eta_{  c[
a}\,\delta_{  b]}^{
d}\,,\qquad R^{  a  b}=-\frac{1}{R^2}\,e^{  a}\,e^{  b}\,,
\ee
where ${R}$ is the $AdS_2$ radius, and the $S^2$ curvature is
\be\label{s2c}
R_{\hat a\hat b\hat c}{}^{\hat d}=-\frac{2}{R^2}\,\eta_{\hat c[\hat
a}\,\delta_{\hat b]}^{\hat
d}\,,\qquad R^{\hat a\hat b}=\frac{1}{R^2}\,e^{\hat a}\,e^{\hat b}\,.
\ee
The $D=4$ gamma--matrices in $AdS_2\times S^2$ are
\be\label{gammaa}
\{\gamma^{\underline a},\gamma^{\underline b}\}=2\,\eta^{\underline{ab}}\,,
\qquad \eta^{\underline {ab}}={\rm diag}\,(-,+,+,+)\,, \qquad \underline a=(a,\hat a)
\ee
\be\label{gamma5}
\gamma^5=i\gamma^0\,\gamma^1\,\gamma^2\,\gamma^3, \qquad
\gamma^5\,\gamma^5=1\,.
\ee
The charge conjugation matrix $C$ is antisymmetric, the matrices $(\gamma^{\underline
a})_{\alpha\beta}\equiv (C\,\gamma^{\underline a})_{\alpha\beta}$ and $(\gamma^{\underline{ab}})_{\alpha\beta}\equiv(C\,\gamma^{\underline{ab}})_{\alpha\beta}$ are symmetric and
$\gamma^5_{\alpha\beta}\equiv (C\gamma^5)_{\alpha\beta}$ is antisymmetric, with
$\alpha,\beta=1,2,3,4$ being the indices of a 4--dimensional spinor representation of $SO(1,3)$.
These $4\times 4$ matrices can be represented in terms
of $2\times 2$ $AdS_2$ gamma--matrices $\rho^a$ ($a=0,1$) and the matrices
 $\rho^{\hat a}$  ($\rho^2=\sigma^1$, $\rho^3=\sigma^3$) associated with $S^2$ as
\be
\gamma^a=\rho^a\otimes \mathbf 1,\qquad \gamma^{\hat a}=\gamma\otimes \rho^{\hat a}\,,\qquad \gamma=\rho^0\rho^1\,.
\ee
$8\times 8$ gamma--matrices associated with $T^6$ are
\be\label{gammaa'}
\{\gamma^{a'},\gamma^{b'}\}=2\,\delta^{{a'}{b'}}\,,\qquad \delta^{a'b'}={\rm
diag}\,(+,+,+,+,+,+)\,,
\ee
\be\label{gamma7}
\gamma^7={i\over{6!}}\,\varepsilon_{\,a_1'a_2'a_3'a_4'a_5'a_6'}\,\gamma^{a_1'}\cdots
\gamma^{a_6'} \qquad
\gamma^7\,\gamma^7=1\,.
\ee
The charge conjugation matrix $C'$ is symmetric and the matrices
$(\gamma^{a'})_{\alpha'\beta'}\equiv (C'\,\gamma^{a'})_{\alpha'\beta'}$ and
$(\gamma^{a'b'})_{\alpha'\beta'}\equiv(C'\,\gamma^{a'b'})_{\alpha'\beta'}$ are
antisymmetric, with
$\alpha',\beta'=1,\cdots,8$ being the indices of an 8--dimensional spinor
representation of
$SO(6)$.

Using the matrices \eqref{gammaa} and \eqref{gammaa'} one can construct the $D=10$
gamma--matrices $\Gamma^A$ as
\bee\label{Gamma10}
&\{\Gamma^A,\,\Gamma^B\}=2\eta^{AB},\qquad
\Gamma^{A}=(\Gamma^{\underline a},\,\Gamma^{a'})\,,\\
&\Gamma^{\underline a}=\gamma^{\underline a}\,\otimes\,{\bf 1},\qquad
\Gamma^{a'}=\gamma^5\,\otimes\,\gamma^{a'},\qquad
\Gamma^{11}=\gamma^5\,\otimes\,\gamma^7,\qquad \underline a=0,1,2,3;\quad
a'=4,\cdots,9\,. \nonumber
\eee
The charge conjugation matrix is
${\mathcal C}=C\otimes C'$. In this realization the
 32--component spinor $\Theta^{\alpha\alpha'}$ is labelled by the
 4--component $AdS_2\times S^2$ spinor index $\alpha$ and the 8--component
 $T^6$ spinor index $\alpha'$.

Finally we can introduce a spinor projection operator which projects onto an 8-dimensional subspace of the 32-dimensional space of spinors as follows
\begin{equation}\label{P8}
\mathcal P_8=\frac{1}{8}(2-iJ_{a'b'}\Gamma^{a'b'}\gamma^7)\,,
\end{equation}
where $J_{a'b'}$ is the K\"ahler form on $T^6$.

\def \rT{{\rm T}}

\def\theequation{B.\arabic{equation}}
\section{Appendix B. Enlarged \psu superalgebra}\label{psualgebra}
\setcounter{equation}0

The \psu superalgebra has the following conventional form. Its bosonic
$SO(2,1)\times SO(3)$ subalgebra is generated by translations $P_{\underline
a}=(P_a, P_{\hat a})$  and $SO(1,1)\times SO(2)$ rotations $M_{\underline{ab}}=(M_{ab},M_{\hat a\hat b})$ in $AdS_2\times S^2$
  \begin{eqnarray}\label{so}
[P_{\underline a}, P_{\underline b}]=-\frac{1}{2}R_{{\underline a}{\underline
b}}{}^{{\underline c}{\underline d}}\,M_{{\underline c}{\underline
d}}\,,\qquad[M_{{\underline a}{\underline b}},P_{\underline
c}]=\eta_{{\underline a}{\underline c}}P_{\underline b}-\eta_{{\underline
b}{\underline c}}P_{\underline a},\nonumber\\
{}[M_{{\underline a}{\underline b}}, M_{{\underline c}{\underline
d}}]=\eta_{{\underline a}{\underline c}}M_{{\underline b}{\underline
d}}+\eta_{{\underline b}{\underline d}}M_{{\underline a}{\underline
c}}-\eta_{{\underline b}{\underline c}}M_{{\underline a}{\underline
d}}-\eta_{{\underline a}{\underline d}}M_{{\underline b}{\underline c}}\,,
\end{eqnarray}
where
\begin{equation}\label{Rua}
R_{\underline{ab}}{}^{\underline{cd}}=(R_{ab}{}^{cd},R_{\hat a\hat b}{}^{\hat
c\hat d})=(\frac{2}{R^2}\delta_{[a}^c\delta_{b]}^d,-\frac{2}{R^2}\delta_{[\hat
a}^{\hat c}\delta_{\hat b]}^{\hat d})
\end{equation}
is the $AdS_2\times S^2$ curvature. The fermionic part of $PSU(1,1|2)$ is
generated by eight Grassmann--odd operators $Q_{\alpha I}$ carrying the index
$\alpha=1,2,3,4$ of a spinorial representation of $SO(1,1)\times SO(2)$ and
the vector index $I=1,2$ of the group $SO(2)$ of external automorphisms of
$PSU(1,1|2)$.\foot{This group should not be confused with the $SO(2)$ rotations of
$S^2$ which is part of $PSU(1,1|2)$.} The generators $Q$ satisfy the following
(anti)commutation relations
\be\label{PSUQ}
[P_{\underline a},Q_I]=\frac{1}{2R}\varepsilon_{IJ}\,Q_J\gamma\gamma_{\underline a}\,,
\qquad [M_{{\underline a}{\underline b}},Q]=-\frac{1}{2}Q\gamma_{{\underline
a}{\underline b}}\,,\qquad [\rT,Q_I]=\frac{1}{2}\,\varepsilon_{IJ}\,Q_J\,,
\ee
\be\label{QQ}
\{Q_I,Q_J\}=
2i\,\delta_{IJ}\,\gamma^{\underline a}\,P_{\underline a}
+\frac{iR}{2}\varepsilon_{IJ}\,\gamma^{{\underline a}{\underline
b}}\,\gamma\,R_{{\underline a}{\underline b}}{}^{{\underline c}{\underline
d}}M_{{\underline c}{\underline d}}\,,
\ee
where the spinorial indices are suppressed, $\gamma^{\underline a}$ and
$\gamma=\gamma^0\gamma^1$ are the $D=4$ gamma matrices with $AdS_2\times S^2$
indices and $\rT$ is the generator of the external $SO(2)$ automorphism of
$PSU(1,1|2)$.
As we will see below,
from  a ten-dimensional point of view it is natural
 to include $\rT$,  even though it is not an element of $\mathfrak{psu}(1,1|2)$.

Since the GS formulation of the string in $AdS_2\times
S^2\times T^6$ uses,  a priori,  $D=10$ notation and 32--component spinors, we
have found it convenient
 to formally lift the \psu
superalgebra to ten dimensions and enlarge it with additional generators
$P_{a'}$ and $M_{a'b'}$ of the U(1) isometries and $SO(6)$ rotations in $T^6$. The
\psu superalgebra enlarged in such a way looks very similar to the
$OSp(6|4)$ superalgebra in the form used in \cite{Sorokin:2010wn}. This similarity allows us to use the same
gamma--matrix identities as those found in \cite{Gomis:2008jt} for the
$AdS_4\times CP^3$ case.

To rewrite the \psu superalgebra in a formally $D=10$ covariant form
we replace the $D=4$ gamma--matrices with their $D=10$ counterparts $\Gamma^A$
and represent the 8 fermionic generators $Q_{\alpha I}$ as 32--component
spinors subject to the 8--dimensional projection
\begin{equation}\label{qpq}
Q=\mathcal P_8\,Q\ ,
\end{equation}
with  $\mathcal P_8$ having 8 non--zero eigenvalues and satisfying
the commutation relations \eqref{PG} and \eqref{PTP}.

The following reasoning demonstrates the relation between the $\mathcal P_8$--projected
32--component spinor $Q_{\alpha\alpha'}$ with the two 4--component $AdS_2\times S^2$
spinors $Q_{\alpha I}$. The projector $\mathcal{P}_8$ and $\Gamma _{(3)}$,
introduced in  eq. \eqref{G(3)}, commute and can be simultaneously
 diagonalized. Because the projector $\mathcal{P}_8$ also commutes
 with the 6d chirality $ \gamma^7$, while $\Gamma _{(3)}$ anti--commutes with
 it, the eigenvalues of $\mathcal{P}_8$ (in the 6d spinor space) are doubly
 degenerate, each pair having opposite values of $\Gamma _{(3)}$. On the other
 hand, $\mathcal{P}_8$ and $\Gamma _{(3)}$ commute with $\Gamma ^{\underline{a}}$
 and thus leave the 4d spinor index $\alpha$ intact. Thus, under the action of
 $\mathcal P_8$, the 6d spinor index $\alpha'$ then reduces to the binary index
  $I=1,2$ that labels the eigenvalues of $\Gamma _{(3)}$ and is associated with
   the $SO(2)$--automorphism.

From the $D=10$ perspective the external $SO(2)$
automorphism of \psu becomes an Abelian subgroup of the group $SO(6)$
of rotations of the $T^6$--torus associated with the K\"ahler form
$J_{a'b'}$ on $T^6$. Namely, the $SO(2)$ generator
$\rT=\frac{1}{6}J^{a'b'}M_{a'b'}$  can be regarded as the part of $M_{a'b'}$ which
acts non--trivially on $Q$. Rewriting the commutator $[T,Q]$ of eq.
\eqref{PSUQ} in ten--dimensional notation we have
\be\label{TQ1}
[\rT,Q]=-\frac{i}{2}Q\gamma^7\mathcal P_8\qquad \Rightarrow \qquad
[M_{a'b'},Q]:=-\frac{i}{2}J_{a'b'} Q\gamma^7\mathcal
P_8=-\frac{1}{2}Q\Gamma_{a'b'}\mathcal P_8,
\ee
where
\begin{equation}\label{g7}
\gamma^7=i\Gamma^{4}\cdots\Gamma^9
\end{equation}
 is the product of the six
gamma--matrices with $T^6$ indices. In \eqref{TQ1} we have used the
gamma--matrix identities
$$\mathcal P_8\Gamma_{a'b'}\mathcal P_8=iJ^{a'b'}\gamma^7\mathcal P_8,\qquad
\gamma^7\mathcal P_8=\mathcal P_8\gamma^7 \qquad {\rm and}\qquad \mathcal
P_8\gamma^7\mathcal P_8:=i\varepsilon\otimes \mathbf{1.}$$
The last expression relates the $2\times 2$ antisymmetric matrix $\varepsilon$
of eqs. \eqref{PSUQ} and \eqref{QQ} times the unit $4\times 4$ spinor matrix in $AdS_2\times S^2$ with the
$\mathcal P_8$--projected matrix $\gamma^7$, eq. \eqref{g7}.

 Now, let us note that $Q$ commutes with the generators $P_{a'}$ of the $U(1)$
isometries of $T^6$. Then, since $\mathcal P_8\Gamma_{a'}\mathcal P_8\equiv
0$, the zero commutator $[P_{a'},Q]$ can be written as
$$[P_{a'},Q]=\frac{i}{2R}Q\gamma\gamma^7\Gamma_{a'}\mathcal P_8\equiv 0,$$
where now $\gamma=\Gamma^0\Gamma^1=\gamma^{01}\otimes\mathbf 1$ is the product of $D=10$ gamma--matrices
carrying $AdS_2$ indices.
Taking all this into account we can write the enlarged \psu superalgebra
in the following form
\begin{eqnarray}
&&[P_A, P_B]=-\frac{1}{2}R_{AB}{}^{CD}M_{CD}\,,\qquad[M_{AB},P_C]=\eta_{AC}P_B-\eta_{BC}P_A\,,\nonumber\\
&&[M_{AB},M_{CD}]=\eta_{AC}M_{BD}+\eta_{BD}M_{AC}-\eta_{BC}M_{AD}-\eta_{AD}M_{BC}\,,\\
&&[P_A,Q]=\frac{i}{2R}Q\gamma\gamma^7\Gamma_A\mathcal P_8\,,\qquad[M_{AB},Q]=-\frac{1}{2}Q\Gamma_{AB}\mathcal P_8\,,
\nonumber\\
&&\{Q,Q\}= 2i(\mathcal P_8\Gamma^A\mathcal P_8)P_A
+\frac{R}{2}(\mathcal P_8\Gamma^{AB}\gamma\gamma^7\mathcal
P_8)R_{AB}{}^{CD}M_{CD}\,,
\label{eq:PSUQ}
\end{eqnarray}
where $M_{AB}=(M_{ab},M_{\hat a\hat b},M_{a'b'})$ and $P_A=(P_a,P_{\hat a},
P_{a'})$ are the generators of Lorentz-trans\-for\-ma\-tions and translations in
$AdS_2$, $S^2$ and $T^6$, respectively, and
\begin{equation}
R_{AB}{}^{CD}=(R_{ab}{}^{cd},R_{\hat a\hat b}{}^{\hat c\hat
d},0)=(\frac{2}{R^2}\delta_{[a}^c\delta_{b]}^d,-\frac{2}{R^2}\delta_{[\hat
a}^{\hat c}\delta_{\hat b]}^{\hat d},0)
\end{equation}
is the curvature tensor of $AdS_2\times S^2\times T^6$.

To recapitulate, the above form of the \psu superalgebra is convenient
for our purposes since it is formulated in a $D=10$  covariant way
and  takes a form very similar to that of $OSp(6|4)$ \cite{Gomis:2008jt,Sorokin:2010wn}.

\def\theequation{C.\arabic{equation}}
\section{Appendix C. Basic relations for the Killing vectors on symmetric
spaces $G/H$}\label{AB}
\setcounter{equation}0
Let $K_M(X)$ or $K_A(X)=e_A{}^M(X)\,K_M(X)$ be the Killing vectors of a
$D$--dimensional symmetric
space $G/H$, where $M$ are world indices and $A$ are tangent space indices.
The Killing vectors
$K_M(X)$ take values in the algebra of the isometry group $G$ and the
one--forms $K=dX^M\,K_M$
satisfy the Maurer--Cartan equations
\begin{eqnarray}\label{K}
dK=-2K\wedge K\ ,\qquad dK\wedge K= K\wedge dK=-2K\wedge K \wedge K\ .
\end{eqnarray}
The following relations also hold
\bee\label{K1}
&[\nabla_A,\nabla_B]K_C=-R_{ABC}{}^D\,K_D\,,\qquad
\nabla_AK_B=[K_A,K_B],\qquad&\no \\
&\nabla_A\nabla_BK_C=[\nabla_AK_B,\,K_C]+[K_B,\nabla_AK_C]=
[\nabla_AK_B,\,K_C]-[\nabla_AK_C,\,K_B]=-2R_{A[BC]}{}^DK_D,\,\,\,\,\,\,\,~&\no \\
&\,[\nabla_AK_B,\,K_C]=[[K_A,\,K_B],\,K_C]=-R_{ABC}{}^D\,K_D,&\no\\
&\Big[[K_A,K_B],[K_C,K_D]\Big]=R_{AB[C}{}^F\,[K_{D]},K_F]-R_{CD[A}{}^F\,[K_{B]},K_F]\,,
 \no \eee
where $R_{ABC}{}^D$ is the curvature of the symmetric space $G/H$.

\def\theequation{D.\arabic{equation}}
\section{Appendix D. Equations of motion from flatness\\ of the Lax connection}\label{AD}
\setcounter{equation}0

We have seen that the Lax connection defined in (\ref{Lax})--(\ref{LF}) is indeed flat for the type IIA and IIB superstring on $AdS_2\times S^2\times T^6$ with RR--flux, provided that the equations of motion are satisfied. To have integrability the opposite should also hold, i.e. requiring flatness of the Lax connection should imply the equations of motion for the string. Here we will show that this is indeed the case.

Computing the curvature of the Lax--connection defined in (\ref{Lax})--(\ref{LF}), (\ref{spectral}) and (\ref{coef}) \emph{without} using the equations of motion we get
\begin{eqnarray}\label{curva}
&&dL-LL=\alpha_2(\nabla\ast J^A-e^D\ast J^{BC}R_{BCD}{}^A)K_A
-\frac{i}{2}\alpha_2(1+\alpha_1)\Theta\Gamma^{AB}\slashed e\mathcal D\Theta\,\nabla_AK_B
\nonumber\\
&&\ \ \ \ \
+(-1)^\nu\frac{i}{2}\alpha_2^2\Theta\Gamma^{AB}\hat\Gamma\slashed e\mathcal D\Theta\,\nabla_AK_B
-\frac{i}{2R}\alpha_2\beta_2\,\Xi\slashed e\mathcal D\Theta
-(-1)^\nu\frac{i}{2R}\alpha_2\beta_1\,\Xi\hat\Gamma\slashed e\mathcal D\Theta\,,
\end{eqnarray}
where $\nu=1(0)$ for type IIA(B) and
\begin{equation}
\slashed e\mathcal D\Theta=\ast e^A\,\Gamma_A\mathcal D\Theta-e^A\,\Gamma_A\hat\Gamma\mathcal D\Theta
\end{equation}
is the quantity that is set to zero by the fermionic equations of motion.

Since the coefficients are all independent (they are different functions of the spectral parameter) each term has to vanish separately. The vanishing of the first term implies
\begin{equation}
\nabla(
\ast e^A+i\Theta\Gamma^A\ast E+i\Theta\Gamma^A\hat\Gamma E
-\frac{i}{8}\ast e^B\,\Theta\Gamma^A\slashed F\Gamma_B\Theta
-\frac{i}{8}e^B\,\Theta\Gamma^A\hat\Gamma\slashed F\Gamma_B\Theta)
-e^D\ast J^{BC}R_{BCD}{}^A=0\,,
\end{equation}
which, using (\ref{JAB}), is equal to the bosonic equations of motion (\ref{bosoneq}).

The vanishing of the last two terms in the curvature implies the $\mathcal P_8$ projection of the fermionic equations of motion,
\begin{equation}\label{P8fe}
\mathcal P_8\slashed e\mathcal D\Theta=0\,.
\end{equation}
We thus get eight fermionic equations of motion associated with the supercoset fermions $\vartheta$.

To single out the remaining 24 fermionic equations we notice that  there are two other terms in \eqref{curva} which have to vanish:
\begin{equation}
\Theta\Gamma^{\underline{ab}}\slashed e\mathcal D\Theta
\end{equation}
and
\begin{equation}
\Theta\Gamma^{\underline{ab}}\hat\Gamma\slashed e\mathcal D\Theta\,,
\end{equation}
where $(\underline{ab})=(01),(23)$ (note that the covariant derivative of $T^6$--translation
 Killing vectors $\nabla_{a'}K_{b'}$ vanishes so the corresponding terms are absent). Using
 the $\mathcal P_8$ projection of the fermionic equation of motion, which we already found,
 we get the conditions
\begin{equation}\label{Lax-conditions}
\upsilon\Gamma^{01}(1-\mathcal P_8)\slashed e\mathcal D\Theta=\upsilon\Gamma^{23}(1-\mathcal
P_8)\slashed e\mathcal D\Theta=\upsilon\Gamma^{01}(1-\mathcal
 P_8)\hat\Gamma\slashed e\mathcal D\Theta=\upsilon\Gamma^{23}(1-\mathcal P_8)\hat\Gamma\slashed e\mathcal D\Theta=0\,.
\end{equation}
This does not directly imply the missing 24 fermionic equations of motion. Instead we get
\begin{equation}\label{feM}
*((1-\mathcal P_8)\slashed e\mathcal D\Theta)=M\upsilon\,,
\end{equation}
where the matrix $M$ has to be such that the four terms
in (\ref{Lax-conditions}) vanish. It is easy to see that there
are non-zero $M$ which satisfy this condition. For example,
 we could take, in the type IIA case, $M\sim\gamma\Gamma_{a'}$
 or $M\sim\Gamma_{a\hat b}\Gamma_{a'b'c'}$. It would seem from
 this analysis that the flatness of the Lax connection does not imply
  the equations of motion for the non-coset fermions but only the weaker
   condition (\ref{feM}). There is,  however,  one more condition that we have yet to
   impose. We can write the equation (\ref{feM}) together with (\ref{P8fe}) as
\begin{equation}\label{feM2}
*(\slashed e\mathcal D\Theta)=(1-\mathcal P_8)M\upsilon\,.
\end{equation}
The left-hand-side of this equation is annihilated by the
 projection matrix $\frac{1}{2}(1+\Gamma)$ with
\begin{equation}\label{Gamma}
\Gamma=\frac{1}{2\sqrt{-h}}\varepsilon^{ij}e_i{}^Ae_j{}^B\Gamma_{AB}\Gamma_{11}\,,
\end{equation}
where $h_{ij}$ is the induced worldsheet metric (when $M=0$ this is simply the statement of kappa-symmetry).
 Therefore, we find the condition
\begin{equation}
(1+\Gamma)(1-\mathcal P_8)M\upsilon=0\,.
\end{equation}
From (\ref{feM}) we also have $\mathcal P_8M=0$. These two conditions imply that
\begin{equation}
0=[1+\Gamma,\mathcal P_8]M\upsilon=[\Gamma,\mathcal P_8]M\upsilon\,.
\end{equation}
From the expressions for $\Gamma$ and $\mathcal P_8$ in (\ref{Gamma}) and (\ref{P8}) it follows that for generic motions of the string the commutator $[\Gamma,\mathcal P_8]$ is non--degenerate and this equation therefore implies that $M=0$. When this is the case eq. (\ref{feM2}) reduces to the fermionic
equations of motion of the string, (\ref{fe}).

There are certain ``singular'' configurations (e.g.,
 when the string moves only in the $AdS_2\times S^2$ subspace with no
  motion along the $T^6$-directions), for which  $[\Gamma,\mathcal P_8]=0$
   as a consequence of the form of $\Gamma$ in (\ref{Gamma}) and the fact that $\mathcal P_8$ involves only $T^6$ gamma--matrices. For these ``singular''
     motions of the string we can not use the above argument to conclude that $M=0$.
      However, we do not expect this kind of a singularity in the Lax connection for
       certain motions of the string, \emph{i.e.} the limit taken to get a ``singular"
        solution from the generic one should be smooth. So the natural conclusion is that
	$M$ should be zero regardless of the solution of the bosonic equations of motion.
	This is also supported by the fact that if we take into consideration the Noether
	 currents associated with rotational isometries in $T^6$, the conservation of these
	  currents will require the matrix $M$ in (\ref{feM2}) to vanish. Unfortunately, it
	  seems that it is not possible to include these currents directly into the Lax
	  connection, since their presence breaks the flatness of the latter.
This subtlety did not appear in the $AdS_4\times CP^3$ case considered in \cite{Sorokin:2010wn}.

We have therefore shown that the string equations of motion indeed
follow from the flatness of the proposed Lax connection (to quadratic
order in fermions) although there is a subtlety with certain ``singular''
classical string solutions.


\providecommand{\href}[2]{#2}\begingroup\raggedright\endgroup

\end{document}